\documentclass[pdflatex,sn-mathphys-num]{sn-jnl}

\usepackage{graphicx}
\usepackage{multirow}
\usepackage{amsmath,amssymb,amsfonts}
\usepackage{amsthm}
\usepackage{mathrsfs}
\usepackage[title]{appendix}
\usepackage{xcolor}
\usepackage{textcomp}
\usepackage{manyfoot}
\usepackage{booktabs}
\usepackage{algorithm}
\usepackage{algorithmicx}
\usepackage{algpseudocode}
\usepackage{listings}

\usepackage{wrapfig}
\usepackage{tabularx}
\usepackage{mathtools}
\usepackage{bm}
\usepackage{siunitx}
\usepackage{cleveref}
\crefname{figure}{Fig.}{Figs.}
\Crefname{figure}{Fig.}{Figs.}
\crefname{equation}{Eq.}{Eqs.}
\Crefname{equation}{Eq.}{Eqs.}
\crefname{table}{Table}{Tables}
\Crefname{table}{Table}{Tables}
\crefname{section}{Sect.}{Sects.}
\Crefname{section}{Sect.}{Sects.}
\crefname{appendix}{Appendix}{Appendices}
\Crefname{appendix}{Appendix}{Appendices}

\crefname{equation}{Eq.}{Eqs.}
\crefname{figure}{Figure}{Figures}
\crefname{section}{Sec.}{Secs.}
\crefname{subsection}{Subsec.}{Subsecs.}
\crefname{table}{Table}{Tables}
\crefname{chapter}{Chapter}{Chapters}
\crefname{part}{Part}{Parts}
\crefname{problem}{Problem}{Problems}
\Crefname{equation}{Equation}{Equations}

\definecolor{MyDarkGreen}{rgb}{0,0.8,0}
\definecolor{MyDarkRed}{rgb}{0.8,0.05,0}
\definecolor{MyDarkBlue}{rgb}{0.0,0.1,0.8}

\newcommand{\sbelDefGradient}{\ensuremath{\mathbf{F}}}

\newcommand{\sbelRightCauchyGreenTensor}{\ensuremath{\mathbf{C}}}
\newcommand{\sbelGreenLagrangeStrainTensor}{\ensuremath{\mathbf{E}}}

\newcommand{\sbelRateOfDeformation}{\ensuremath{\mathbf{D}}}

\newcommand{\sbelInfVolumeRef}{\ensuremath{dV_r}}
\newcommand{\sbelVolumeRef}{\ensuremath{V_r}}

\newcommand{\sbelVolumeParent}{\ensuremath{V_p}}
\newcommand{\sbelInfVolumeParent}{\ensuremath{dV_p}}

\newcommand{\sbelStrainEnergyDensityFunction}{\ensuremath{\psi}}
\newcommand{\sbelDampingEnergyDensityFunction}{\ensuremath{\phi}}

\newcommand{\sbelNUmatrix}{\ensuremath{\mathbf{N}}}

\newcommand{\sbelNUSub}[1]{\ensuremath{{\mathbf{e}}_{#1}}}

\newcommand{\sbelNUcount}{\ensuremath{n_u}} 

\newcommand{\sbelShapeFunArray}{\ensuremath{\mathbf{s}}}
\newcommand{\sbelShapeFun}{\ensuremath{s}}
\newcommand{\sbelShapeFunJac}{\ensuremath{\mathbf{H}}}
\newcommand{\sbelShapeFunElementGrad}{\ensuremath{\mathbf{h}}}
\newcommand{\sbelShapeFunArrayParent}{\ensuremath{{\mathbf{s}}^p}}
\newcommand{\sbelShapeFunParent}{\ensuremath{s^p}}
\newcommand{\sbelShapeFunJacParent}{\ensuremath{\mathbf{H}^p}}

\newcommand{\sbelPoint}[1]{\textbf{#1}}

\newcommand{\sbelBodySub}[1]{\ensuremath{\mathcal{B}_{#1}}}

\newcommand{\sbelFirstPiolaKirchhoffStress}{\ensuremath{\mathbf{P}}}

\newcommand{\sbelSecondPiolaKirchhoffStress}{\ensuremath{\mathbf{S}}}

\definecolor{arsenic}{rgb}{0.23, 0.27, 0.29}
\definecolor{charcoal}{rgb}{0.21, 0.27, 0.31}
\definecolor{hanblue}{rgb}{0.27, 0.42, 0.81}
\definecolor{blue-ncs}{rgb}{0.0, 0.53, 0.74}
\definecolor{awesome}{rgb}{1.0, 0.13,0.32}
\definecolor{darkgreen}{rgb}{0, .4,0}

\crefname{equation}{Eq.}{Eqs.}       
\Crefname{equation}{Equation}{Equations}  

\providecommand{\sbelNUmatrix}{\mathbf{N}}
\providecommand{\sbelPoint}[1]{\ifmmode\mathsf{#1}\else\textsf{#1}\fi}
\providecommand{\sbelBodySub}[1]{\ifmmode\mathrm{#1}\else\textrm{#1}\fi}

\newcolumntype{L}[1]{>{\raggedright\arraybackslash}p{#1}}
\newcolumntype{C}[1]{>{\centering\arraybackslash}p{#1}}
\newcolumntype{Y}{>{\raggedright\arraybackslash}X}

\theoremstyle{thmstyletwo}

\graphicspath{{images/}{/home/jason/Desktop/local-image-archive/journals/2026/tl-fea/}{/home/zzhou292/Desktop/STUDY/local-image-archive/journals/2026/tl-fea/}{../../../../local-image-archive/journals/2026/tl-fea/}{../../../../../local-image-archive/journals/2026/tl-fea/}}
\newcommand{\safeincludegraphics}[2][]{%
  \IfFileExists{images/#2}{%
    \includegraphics[#1]{images/#2}%
  }{%
    \IfFileExists{/home/jason/Desktop/local-image-archive/journals/2026/tl-fea/#2}{%
      \includegraphics[#1]{/home/jason/Desktop/local-image-archive/journals/2026/tl-fea/#2}%
    }{%
      \IfFileExists{/home/zzhou292/Desktop/STUDY/local-image-archive/journals/2026/tl-fea/#2}{%
        \includegraphics[#1]{/home/zzhou292/Desktop/STUDY/local-image-archive/journals/2026/tl-fea/#2}%
      }{%
        \IfFileExists{../../../../local-image-archive/journals/2026/tl-fea/#2}{%
          \includegraphics[#1]{../../../../local-image-archive/journals/2026/tl-fea/#2}%
        }{%
          \IfFileExists{../../../../../local-image-archive/journals/2026/tl-fea/#2}{%
            \includegraphics[#1]{../../../../../local-image-archive/journals/2026/tl-fea/#2}%
          }{%
            \begingroup
            \setlength{\fboxsep}{8pt}%
            \fbox{%
              \parbox[c][0.16\textheight][c]{0.82\linewidth}{%
                \centering
                \textbf{Missing figure}\\[0.5ex]
                \small\texttt{\detokenize{#2}}%
              }%
            }%
            \endgroup
          }%
        }%
      }%
    }%
  }%
}

\begin{document}

\title[TL-FEA Part I: Formulation]{A Total Lagrangian Finite Element Framework
for Multibody Dynamics: Part I -- Formulation}

\author*[1]{\fnm{Zhenhao} \sur{Zhou}}\email{zzhou292@wisc.edu}

\author[1]{\fnm{Ganesh} \sur{Arivoli}}\email{arivoli@wisc.edu}

\author[1]{\fnm{Dan} \sur{Negrut}}\email{negrut@wisc.edu}

\affil*[1]{\orgdiv{Department of Mechanical Engineering},
  \orgname{University of Wisconsin--Madison},
  \orgaddress{\city{Madison}, \state{WI}, \postcode{53706}, \country{USA}}}

\abstract{We present a Total Lagrangian finite element framework for
finite-deformation multibody dynamics. The framework combines a compact
kinematic representation, a deformation-gradient-based formulation, an
element-agnostic constitutive interface, and a systematic
constraint-construction machinery for coupling deformable bodies through
engineering joints. Within this setting, we derive the equations of motion
for collections of deformable bodies, and formulate their response in the
presence of external loads, frictional contact forces, and constraint
reaction forces. The framework accommodates field forces applied pointwise,
over surfaces, or throughout volumes, and supports material models of
practical interest, including Mooney--Rivlin, Neo-Hookean, and
Kelvin--Voigt. A companion paper discusses the GPU-accelerated
implementation of the framework outlined herein, and reports on numerical
experiments and benchmark results.}


\let\origprintkeywords\printkeywords
\renewcommand{\printkeywords}{}
\maketitle
\renewcommand{\printkeywords}{\origprintkeywords}

\noindent\textbf{Article Highlights}
\begin{itemize}
  \item A deformation-gradient-based Total Lagrangian FEA framework for finite-deformation flexible multibody dynamics.
  \item An element-agnostic constitutive interface supports Mooney--Rivlin, Neo-Hookean, and Kelvin--Voigt models.
  \item Engineering joints are assembled from constraint primitives, enabling systematic coupling of deformable bodies.
\end{itemize}

\origprintkeywords

\section{Introduction}
\label{sec:introduction}
This paper presents a Total Lagrangian finite element formulation for constrained multibody dynamics with finite deformations. The focus is on the formulation itself: kinematics, constitutive response, external loading, frictional contact, and bilateral constraints are written in a common notation that is compatible with implicit time integration. The numerical methods, GPU implementation, and benchmark problems built on this formulation are addressed in the companion paper~\cite{json-ruochun-ganesh-danTLFEA-2-2026}, which is currently under review at Engineering with Computers.

In the Total Lagrangian setting, all kinematic and stress quantities are referred to the reference configuration. This choice is standard in geometrically nonlinear solid mechanics~\cite{belytschko00,Bonet2016nonlinear} and leads to a clean separation between reference-domain geometry and deformation-dependent response through the deformation gradient $\mathbf{F}$ and derived strain measures. It also allows derivatives of the interpolation to be evaluated with respect to the reference domain, which is attractive in large-deformation settings; see, e.g., \cite{peng2023comparison} for comparisons against the Updated Lagrangian and corotational alternatives.

Within flexible multibody dynamics, the Absolute Nodal Coordinate Formulation (ANCF)~\cite{Shabana1997,shabanaYakoub2001Part1} can be viewed as a particular instance of this broader Total Lagrangian perspective. ANCF uses absolute positions and their spatial gradients as nodal coordinates, which yields a polynomial position field valid under large rotations and deformations, produces a constant mass matrix, and uses the same deformation gradient $\mathbf{F}$ as standard continuum finite elements. A related variant by Zhang~\cite{zhang2021direct} eliminates the deformation gradient entirely in an explicit, single-body setting; the present work retains $\mathbf{F}$ as the central kinematic quantity, since implicit constrained multibody dynamics requires both the tangent stiffness and the constraint Jacobians that derive from it.

A central modeling choice concerns how the nodal unknowns enter the position field. In much of the finite element and ANCF literature, the interpolation is written in the form $\mathbf{r}=\mathbf{S}(\mathbf{u})\,\mathbf{e}(t)$, where $\mathbf{u}$ denotes the reference coordinate, $\mathbf{S}$ is assembled from shape functions, and $\mathbf{e}$ collects the nodal unknowns~\cite{belytschko00,shabana2020}. Here, the same interpolation is written in the equivalent compact form $\mathbf{r}=\mathbf{N}(t)\,\mathbf{s}(\mathbf{u})$, where $\mathbf{N}$ stores the nodal unknowns as columns and $\mathbf{s}$ contains the scalar shape functions. In this notation, the deformation gradient takes the form $\mathbf{F}=\mathbf{N}(t)\,\mathbf{H}(\mathbf{u})$, with $\mathbf{H}$ collecting derivatives of the shape functions with respect to the reference coordinates. Earlier contributions in this direction include the edge-matrix form used for linear tetrahedra~\cite{sifakis2012fem} and the component-free Lagrangian formulation of~\cite{stickle2022component}, which avoids Voigt notation but retains the tensor summation form. The functional value of this representation is that it renders the constraint Jacobian derivation element-agnostic: because $\mathbf{F}$, the internal force gradient, and the first variation of the position field all factor through the same $\mathbf{N}$--$\mathbf{H}$ split, constraint derivatives across all element types follow from a single chain-rule pattern applied to the point-evaluation operator, without element-specific case analysis.

The treatment of kinematic constraints is a central part of the formulation. In rigid-body dynamics, primitive constraints such as DP1, DP2, CD, and DIST~\cite{jayAllieDan2022rA} provide a standard route to the construction of engineering joints. For deformable bodies, Betsch and Steinmann~\cite{betsch2003constrained} developed rotationless constrained dynamics for geometrically exact beams, but their treatment is specific to that element class and does not extend to isoparametric solid or shell elements, nor does it address the Jacobian accumulation and conditioning issues that arise when dot-product and coordinate-difference constraints are mixed. Sugiyama et al.~\cite{sugiyama2003formulation} formulated ANCF joint constraints using position-gradient degrees of freedom as orientation surrogates, providing a foundation for deformable-body joints; however, their derivation is cast in the block shape-function notation of ANCF and does not supply the element-level Jacobian accumulation rules, the curvature contribution to the Newton system, or the row-scaling analysis needed for a well-conditioned mixed constraint set. The present work addresses these gaps by applying the primitive-constraint viewpoint directly to isoparametric finite elements in the $\mathbf{N}(t)\,\mathbf{s}(\mathbf{u})$ setting, deriving explicit element-level Jacobian blocks for each primitive, characterizing the DP1 Hessian structure, and providing a conditioning analysis for mixed CD/DP1 joint systems.

On the constitutive side, the formulation uses the first Piola--Kirchhoff stress $\mathbf{P}(\mathbf{F})$ and its material derivative $\partial\mathbf{P}/\partial\mathbf{F}$ as the interface between the constitutive law and the spatial discretization~\cite{belytschko00,Bonet2008nonlinear,zienkiewicz-taylor-FEA2005}. This separates constitutive modeling from element topology in a natural way. Within this setting, the paper derives nodal internal-force expressions and consistent linearizations for several hyperelastic and viscoelastic material models. In the ANCF literature, Garc\'{i}a-Vallejo et al.~\cite{garcia2004efficient} derived elastic force Jacobians for the St.~Venant--Kirchhoff model, while subsequent work on nonlinear constitutive models~\cite{maqueda2007poisson,Orzechowski2015} relied on numerical tangent approximations. The present work provides closed-form tangents for the Mooney--Rivlin and Kelvin--Voigt models in the same $\mathbf{P}(\mathbf{F})$ interface, making them directly available for any element topology supported by the framework.

The contributions of this paper are as follows. First, it derives a systematic treatment of deformable-body joint constraints for isoparametric finite elements, including explicit element-level Jacobian accumulation rules for all primitive constraint types, the consistent Newton linearization with the DP1 curvature term, and a conditioning analysis for mixed CD/DP1 systems that identifies the source of ill-conditioning and prescribes reference-configuration row weights that restore a balanced Newton system. Second, it derives closed-form internal-force expressions and constitutive tangents for the SVK, Mooney--Rivlin, and Kelvin--Voigt models through a unified $\mathbf{P}(\mathbf{F})$ interface that is independent of element topology. Third, it assembles kinematics, constitutive response, external loading, frictional contact, and bilateral constraints into a single virtual-work framework, providing a consistent derivation of every force and reaction term that enters the equations of motion. Fourth, it recasts the backward-Euler constrained TL-FEA step as a velocity-level augmented-Lagrangian optimization problem: the stationarity conditions of a single scalar objective recover the discrete equations of motion, bilateral constraints are enforced through multiplier and penalty terms whose transpose action is assembled element-by-element without forming a dense global constraint matrix, and the structure of the resulting Newton system --- including the positive semidefinite Gauss--Newton term and the indefinite DP1 curvature correction --- is analyzed explicitly, linking the choice of linearization to solver requirements. Claims regarding computational efficiency, robustness, and large-scale numerical performance are addressed in the companion paper~\cite{json-ruochun-ganesh-danTLFEA-2-2026}.

The remainder of the paper is organized as follows. Section~\ref{sec:kinematic_aspects} introduces the TL deformation map, the $\mathbf{N}(t)\,\mathbf{s}(\mathbf{u})$ interpolation, and the constraint primitives and engineering joints. Section~\ref{sec:mat_models_aspects} presents the stress measures and material models. Section~\ref{sec:dynamics_aspects} derives the virtual-work contributions for inertia, body forces, internal forces, concentrated loads, and surface tractions. Section~\ref{sec:frictional_contact_aspects} describes the frictional contact model. Section~\ref{sec:approximation_of_integrals} covers numerical quadrature in the isoparametric setting. Section~\ref{sec:alm_aspects} formulates the augmented-Lagrangian time-discrete step. Section~\ref{sec:conclusions} summarizes the contributions. Appendices provide shape function data for the element types used in Part~II, details of the joint constraint linearization, and closed-form tangent expressions for all material models.

\section{Kinematics Aspects}
\label{sec:kinematic_aspects}
\subsection{The TL Deformation Map}
\label{subsec:deformation_map}

Large-deformation multibody dynamics can be formulated in a continuum setting by describing, for each element, the deformation map $\mathbf r=\boldsymbol{\phi}(\mathbf u;t)\in\mathbb{R}^3$ and the associated deformation gradient $\mathbf F(\mathbf u;t)=\partial \mathbf r/\partial \mathbf u$, where $\mathbf u=[u,v,w]^T\in\mathbb{R}^3$ denotes a material point in the element reference configuration and $\mathbf r(\mathbf u;t)$ gives its current location at time $t$.

\begin{figure}[t]
  \centering
  \safeincludegraphics[width=0.64\linewidth]{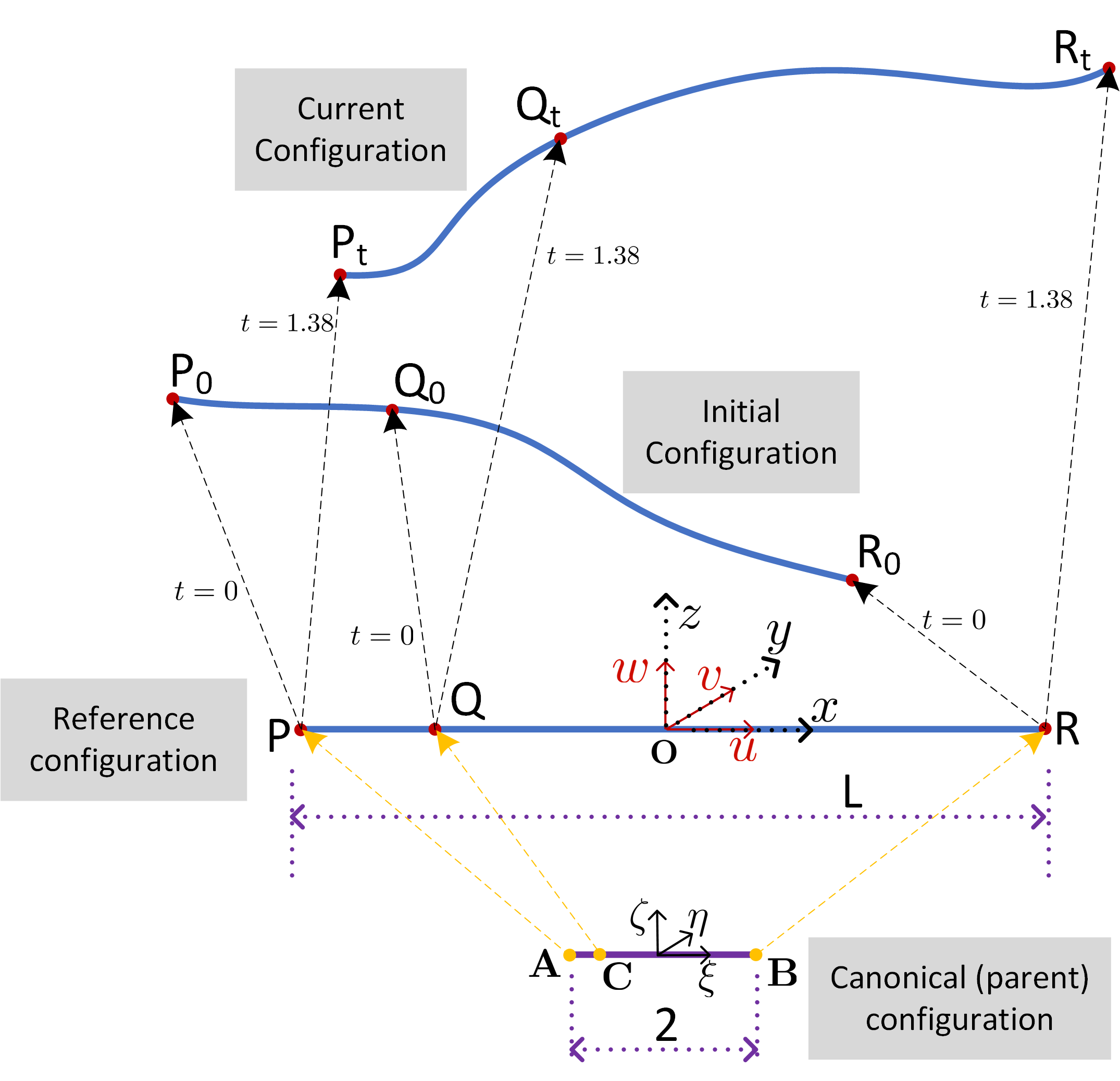}
  \caption{Four configurations associated with one element in TL-FEA. The parent and reference configurations are auxiliary constructs used for interpolation and constitutive description; the initial and current configurations are the physical configurations at $t=0$ and at the current time, respectively}
  \label{fig:element_configurations}
\end{figure}

Fig.~\ref{fig:element_configurations} distinguishes four configurations used throughout the paper. The reference configuration is the stress-free, strain-free configuration of the element, typically supplied by a mesh generator such as \texttt{Gmsh}~\cite{GeuzaineRemacle2009Gmsh} or \texttt{TetGen}~\cite{Si2015TetGen}. The current configuration is described relative to a fixed inertial frame $Oxyz$ through the map $\mathbf r(u,v,w;t)=[x(u,v,w;t),y(u,v,w;t),z(u,v,w;t)]^T$. The initial configuration is the physical configuration at $t=0$, which may already carry stress. Finally, the parent (canonical) configuration is introduced for numerical integration; depending on the element, the parent-to-reference map is either isoparametric or affine.

In standard TL-FEA, the unknowns are the nodal quantities that parameterize the current configuration of the element. For a T10 tetrahedral element~\cite{zienkiewicz-taylor-FEA2005}, these are the ten nodal positions $\mathbf e_1(t),\ldots,\mathbf e_{10}(t)\in\mathbb{R}^3$. Within flexible multibody dynamics, the Absolute Nodal Coordinate Formulation (ANCF)~\cite{Shabana1996} is another TL-FEA instance, distinguished by the use of nodal positions together with spatial gradients as degrees of freedom. The ANCF literature spans beam, cable, shell, and solid-like elements, together with remedies for locking and related pathologies~\cite{berzeri2000,GarciaVallejo2007LockingFree,Gerstmayr2013,Otsuka2022RecentAdvances,Shabana2023Overview}. For present purposes, the relevant commonality is that both classical TL-FEA and ANCF describe a continuum map through the deformation gradient and strain measures referred to the reference configuration.

In the standard notation, the map $\boldsymbol{\phi}$ is approximated by interpolating the nodal unknowns as
\begin{equation}
  \label{eq:r_eval_position_ancf}
  \mathbf r(\mathbf u;t)=\mathbf S(\mathbf u)\:\mathbf e(t) \; ,
\end{equation}
where $\mathbf S(\mathbf u)\in\mathbb{R}^{3\times 3n}$ is a block shape-function matrix and $\mathbf e(t)\in\mathbb{R}^{3n}$ collects the nodal degrees of freedom~\cite{shabanaYakoub2001Part1,Shabana2023Overview}. For instance, for the T10 tetrahedron $n=10$, whereas for a fully parameterized two-node beam element $n=8$.

In this work, the same map is written instead as
\begin{equation}
  \label{eq:r_eval_position}
  \mathbf r(\mathbf u;t)=\mathbf N(t)\:\mathbf s(\mathbf u) \; ,
\end{equation}
where $\mathbf N(t)=[\mathbf e_1(t),\ldots,\mathbf e_n(t)]\in\mathbb{R}^{3\times n}$ stores the nodal unknowns columnwise and $\mathbf s(\mathbf u)\in\mathbb{R}^n$ collects the scalar shape functions. In classical TL-FEA, the columns of $\mathbf N$ are nodal positions; in ANCF, they may include positions together with slope-related quantities. This representation is algebraically equivalent to \cref{eq:r_eval_position_ancf}, but it will be used throughout because it leads to a compact expression for the deformation gradient, virtual displacements, and the Jacobian blocks that enter the constraint and internal-force calculations.

From \cref{eq:r_eval_position}, the velocity, acceleration, and virtual displacement of the current point associated with $\mathbf u$ are
\begin{subequations}
\begin{equation}
  \label{eq:expression-rdot}
  \dot{\mathbf r}(u,v,w;t)=\dot{\sbelNUmatrix}(t)\:\sbelShapeFunArray(u,v,w), \qquad \ddot{\mathbf r}(u,v,w;t)=\ddot{\sbelNUmatrix}(t)\:\sbelShapeFunArray(u,v,w),
\end{equation}
and
\begin{equation}
  \label{eq:eq:firstOrderVariation-r}
  \delta\mathbf r^T=\sbelShapeFunArray^T(u,v,w)\,(\delta\sbelNUmatrix)^T=\sbelShapeFunArray^T(u,v,w)\begin{bmatrix}\delta\sbelNUSub{1}^T\\ \delta\sbelNUSub{2}^T\\ \vdots\\ \delta\sbelNUSub{n}^T\end{bmatrix}\; .
\end{equation}
\end{subequations}

Let $\sbelShapeFun_i(u,v,w)$ denote the $i$-th entry of $\sbelShapeFunArray(u,v,w)$ and define
\begin{subequations}
\begin{equation}
  \label{eq:shape-func-def-grad-comp}
  \sbelShapeFunElementGrad_i(u,v,w)\equiv\nabla_{\mathbf u}\,\sbelShapeFun_i(u,v,w)\in\mathbb{R}^3,
\end{equation}
and
\begin{equation}
  \label{eq:shape-func-def-grad}
  \mathbf H(u,v,w)=\begin{bmatrix}\mathbf h_1^T(u,v,w)\\ \mathbf h_2^T(u,v,w)\\ \vdots\\ \mathbf h_n^T(u,v,w)\end{bmatrix}=\frac{\partial \mathbf s(u,v,w)}{\partial \mathbf u}\in\mathbb{R}^{n\times 3}\; .
\end{equation}
\end{subequations}
The deformation gradient then takes the form
\begin{equation}
  \label{eq:expression-F}
  \mathbf F=\frac{\partial \mathbf r}{\partial \mathbf u}=\mathbf N(t)\:\mathbf H(u,v,w)=\sum_{i=1}^{n}\mathbf e_i(t)\,\mathbf h_i^T(u,v,w)\; .
\end{equation}

The strain measures used in the remainder of the paper are the right Cauchy--Green tensor and the Green--Lagrange strain tensor,
\begin{subequations}
\begin{equation}
  \label{eq:right-cauchy-green-tensor}
  \sbelRightCauchyGreenTensor\coloneqq\mathbf F^T\mathbf F \, ,
\end{equation}
\begin{equation}
  \label{eq:green-lagrange-strain-tensor}
  \mathbf E\coloneqq\frac{1}{2}(\sbelRightCauchyGreenTensor-\mathbf I)=\frac{1}{2}(\mathbf F^T\mathbf F-\mathbf I)\, .
\end{equation}
\end{subequations}
The expressions of $\mathbf s$ and $\mathbf H$ for several element types are provided in Appendix~\ref{sec:select_element_types}.


\subsection{Kinematic Constraints}
\label{subsec:kinematic_constraints}

We consider bilateral holonomic constraints built from four scalar geometric primitives---dot-product~1 (DP1), dot-product~2 (DP2), distance (DIST), and coordinate-difference (CD). These four primitives are sufficient to show how deformable-body joints are expressed in the proposed TL-FEA notation and how the corresponding Jacobian blocks follow from the point-evaluation operator \cref{eq:r_eval_position} and its first variation \cref{eq:eq:firstOrderVariation-r}. Richer primitive libraries can be handled in the same way, but are not needed here.

Each primitive is posed as a scalar position-level condition $c(\mathbf q,t)=0$. Since the discrete step is written as $\mathbf q_{n+1}=\mathbf q_n+h\,\mathbf v$, the nonlinear solve at time $t_{n+1}$ requires the constraint value $c(\mathbf q_{n+1},t_{n+1})$ and its Jacobian $J_c(\mathbf q_{n+1},t_{n+1})=\partial c/\partial \mathbf q$, which supplies the constraint-direction contributions entering the augmented-Lagrangian residual in Eq.~\eqref{eq:residual}. The purpose of this subsection is therefore threefold: to define the primitive constraints, to show how they are composed into standard engineering joints, and to make explicit the Jacobian accumulation and linearization issues that arise once dot-product constraints are present.

The primitives DP1, DP2, DIST, and CD are scalar compositions of point evaluations \cref{eq:r_eval_position} and point differences, so their Jacobians follow the same chain-rule pattern. For compactness, each primitive is stated through its defining scalar equation, its first variation, and the corresponding nonzero element-level Jacobian blocks.

\subsubsection{Dot-product 1 (DP1)}
\label{subsubsec:dp1}

DP1 prescribes the dot product between two directions, each defined by a pair of points. Let $\mathcal P,\mathcal Q$ lie on body $b$ (elements $E,F$) and $\mathcal R,\mathcal T$ lie on body $c$ (elements $G,H$). With the directions
\begin{equation}
\mathbf a=\mathbf r_{\mathcal Q}-\mathbf r_{\mathcal P},\qquad
\mathbf b=\mathbf r_{\mathcal T}-\mathbf r_{\mathcal R},
\label{eq:dp1_ab}
\end{equation}
DP1 enforces
\begin{equation}
c_{\mathrm{DP1}}(\mathbf q,t)=\mathbf a^{\mathsf T}\mathbf b-f(t)=0,
\label{eq:dp1_constraint}
\end{equation}
where $f(t)$ is prescribed (often constant; $f\equiv 0$ yields a perpendicularity condition). The first variation is
\begin{equation}
\delta c_{\mathrm{DP1}}=\mathbf b^{\mathsf T}\delta\mathbf a+\mathbf a^{\mathsf T}\delta\mathbf b,\qquad \delta\mathbf a=\delta\mathbf r_{\mathcal Q}-\delta\mathbf r_{\mathcal P},\quad \delta\mathbf b=\delta\mathbf r_{\mathcal T}-\delta\mathbf r_{\mathcal R}.
\label{eq:dp1_variation}
\end{equation}
Using \cref{eq:eq:firstOrderVariation-r} at each point, the only nonzero Jacobian blocks are those associated with the four hosting elements. For example, the contribution associated with point $\mathcal P$ has the form
\begin{equation}
\frac{\partial c_{\mathrm{DP1}}}{\partial \mathbf e_E^b}
=-\mathbf b^{\mathsf T}\left(\mathbf s_E^b(\mathbf u_{\mathcal P})^{\mathsf T}\otimes \mathbf I_3\right),
\label{eq:dp1_block_P}
\end{equation}
and the blocks associated with $\mathcal Q,\mathcal R,\mathcal T$ follow by replacing $(E,\mathcal P,-\mathbf b)$ with $(F,\mathcal Q,+\mathbf b)$, $(G,\mathcal R,-\mathbf a)$, and $(H,\mathcal T,+\mathbf a)$, respectively.

\subsubsection{Dot-product 2 (DP2)}
\label{subsubsec:dp2}

DP2 prescribes the dot product between a body-attached direction and a connector direction. In DP1, each direction is defined by two points that lie on the same body. In keeping with the nomenclature of~\cite{jayAllieDan2022rA}, DP2 instead uses three points on one body and one point on the other. Let $\mathcal P,\mathcal Q,\mathcal R$ lie on body $b$ (elements $E,F,G$) and $\mathcal T$ lie on body $c$ (element $H$). Define
\begin{equation}
\mathbf a=\mathbf r_{\mathcal Q}-\mathbf r_{\mathcal P},\qquad
\mathbf b=\mathbf r_{\mathcal T}-\mathbf r_{\mathcal R},
\label{eq:dp2_ab}
\end{equation}
where $\mathbf a$ is a body-attached direction on body $b$ and $\mathbf b$ is a connector direction from point $\mathcal R$ on body $b$ to point $\mathcal T$ on body $c$. DP2 enforces
\begin{equation}
c_{\mathrm{DP2}}(\mathbf q,t)=\mathbf a^{\mathsf T}\mathbf b-f(t)=0,
\label{eq:dp2_constraint}
\end{equation}
with first variation
\begin{equation}
\delta c_{\mathrm{DP2}}=\mathbf b^{\mathsf T}\delta\mathbf a+\mathbf a^{\mathsf T}\delta\mathbf b,\qquad \delta\mathbf a=\delta\mathbf r_{\mathcal Q}-\delta\mathbf r_{\mathcal P},\quad \delta\mathbf b=\delta\mathbf r_{\mathcal T}-\delta\mathbf r_{\mathcal R}.
\label{eq:dp2_variation}
\end{equation}
The only nonzero Jacobian blocks are associated with the hosting elements of $\mathcal P,\mathcal Q,\mathcal R$, and $\mathcal T$, and are given by
\begin{align}
\frac{\partial c_{\mathrm{DP2}}}{\partial \mathbf e_E^b}
&=-\mathbf b^{\mathsf T}\left(\mathbf s_E^b(\mathbf u_{\mathcal P})^{\mathsf T}\otimes \mathbf I_3\right),&
\frac{\partial c_{\mathrm{DP2}}}{\partial \mathbf e_F^b}
&=+\mathbf b^{\mathsf T}\left(\mathbf s_F^b(\mathbf u_{\mathcal Q})^{\mathsf T}\otimes \mathbf I_3\right),\nonumber\\
\frac{\partial c_{\mathrm{DP2}}}{\partial \mathbf e_G^b}
&=-\mathbf a^{\mathsf T}\left(\mathbf s_G^b(\mathbf u_{\mathcal R})^{\mathsf T}\otimes \mathbf I_3\right),&
\frac{\partial c_{\mathrm{DP2}}}{\partial \mathbf e_H^c}
&=+\mathbf a^{\mathsf T}\left(\mathbf s_H^c(\mathbf u_{\mathcal T})^{\mathsf T}\otimes \mathbf I_3\right).
\label{eq:dp2_blocks}
\end{align}

\subsubsection{Distance (DIST)}
\label{subsubsec:dist}

DIST prescribes the separation between two points. Let $\mathcal P$ lie on body $b$ (element $E$) and $\mathcal Q$ lie on body $c$ (element $F$), and define $\mathbf a=\mathbf r_{\mathcal Q}-\mathbf r_{\mathcal P}$. We employ the squared-distance form
\begin{equation}
c_{\mathrm{DIST}}(\mathbf q,t)=\frac{1}{2}\left(\mathbf a^{\mathsf T}\mathbf a-f(t)^2\right)=0,
\label{eq:dist_constraint}
\end{equation}
with $f(t)>0$. The first variation is
\begin{equation}
\delta c_{\mathrm{DIST}}=\mathbf a^{\mathsf T}\delta\mathbf a,\qquad \delta\mathbf a=\delta\mathbf r_{\mathcal Q}-\delta\mathbf r_{\mathcal P}.
\label{eq:dist_variation}
\end{equation}
Therefore, the only nonzero Jacobian blocks are those for the hosting elements of $\mathcal P$ and $\mathcal Q$:
\begin{equation}
\frac{\partial c_{\mathrm{DIST}}}{\partial \mathbf e_E^b}
=-\mathbf a^{\mathsf T}\left(\mathbf s_E^b(\mathbf u_{\mathcal P})^{\mathsf T}\otimes \mathbf I_3\right),
\qquad
\frac{\partial c_{\mathrm{DIST}}}{\partial \mathbf e_F^c}
=+\mathbf a^{\mathsf T}\left(\mathbf s_F^c(\mathbf u_{\mathcal Q})^{\mathsf T}\otimes \mathbf I_3\right).
\label{eq:dist_blocks}
\end{equation}

\subsubsection{Coordinate-difference (CD)}
\label{subsubsec:cd}

CD constrains one Cartesian component of a point difference. Let $\mathcal P$ lie on body $b$ (element $E$) and $\mathcal Q$ lie on body $c$ (element $F$), and define $\mathbf a=\mathbf r_{\mathcal Q}-\mathbf r_{\mathcal P}$. Given a constant unit vector $\mathbf d\in\{[1,0,0]^{\mathsf T},[0,1,0]^{\mathsf T},[0,0,1]^{\mathsf T}\}$ selecting the $x$, $y$, or $z$ component, CD enforces
\begin{equation}
c_{\mathrm{CD}}(\mathbf q,t)=\mathbf d^{\mathsf T}\mathbf a-f(t)=0.
\label{eq:cd_constraint}
\end{equation}
The first variation is
\begin{equation}
\delta c_{\mathrm{CD}}=\mathbf d^{\mathsf T}\delta\mathbf a,\qquad \delta\mathbf a=\delta\mathbf r_{\mathcal Q}-\delta\mathbf r_{\mathcal P},
\label{eq:cd_variation}
\end{equation}
so that the only nonzero Jacobian blocks are
\begin{equation}
\frac{\partial c_{\mathrm{CD}}}{\partial \mathbf e_E^b}
=-\mathbf d^{\mathsf T}\left(\mathbf s_E^b(\mathbf u_{\mathcal P})^{\mathsf T}\otimes \mathbf I_3\right),
\qquad
\frac{\partial c_{\mathrm{CD}}}{\partial \mathbf e_F^c}
=+\mathbf d^{\mathsf T}\left(\mathbf s_F^c(\mathbf u_{\mathcal Q})^{\mathsf T}\otimes \mathbf I_3\right).
\label{eq:cd_blocks}
\end{equation}
Vector coincidence constraints are obtained by stacking three CD equations with $\mathbf d=\mathbf e_x,\mathbf e_y,\mathbf e_z$.

\subsubsection{Engineering Joints by Constraint Composition}
\label{subsubsec:joints}

Each scalar primitive contributes one scalar kinematic condition. Engineering joints are obtained by stacking such primitives. The degree-of-freedom counts reported in \cref{tab:joint_decomp} are local counts that hold when the assembled constraint rows are independent. In particular, the point sets used to define the joint must be regular: points used to define directions must remain distinct, the resulting direction vectors must be nonzero, off-axis directions must be linearly independent when required, and the assembled joint Jacobian must have full row rank. Under these assumptions, stacking $m$ independent scalar constraints removes $m$ relative degrees of freedom.

\cref{tab:joint_decomp} catalogs the primitive decomposition for six standard joint types.

\begin{table}[t]
  \centering
  \caption{Primitive decomposition for engineering joints. CD denotes coordinate-difference (point-coincidence) constraints and DP denotes dot-product constraints. For the cylindrical and prismatic joints the notation $a{+}b$ indicates $a$ axis-parallelism (DP1-pattern) plus $b$ offset-collinearity (DP2-pattern) constraints}
  \label{tab:joint_decomp}
  \begin{tabular}{@{}lcccl@{}}
    \toprule
    Joint type   & CD & DP      & Total $m$ & Remaining DOF \\
    \midrule
    Spherical    & 3  & 0       & 3         & 3 rotational \\
    Universal    & 3  & 1       & 4         & 2 rotational \\
    Revolute     & 3  & 2       & 5         & 1 rotational \\
    Fixed (weld) & 3  & 3       & 6         & 0 \\
    Cylindrical  & 0  & $2{+}2$ & 4         & 1 rot.\ + 1 transl. \\
    Prismatic    & 0  & $3{+}2$ & 5         & 1 translational \\
    \bottomrule
  \end{tabular}
\end{table}

The table should therefore be read as a catalog of constraint compositions, not as an unconditional guarantee. Degenerate point choices or near-singular configurations reduce the rank of the assembled Jacobian and change the effective mobility. We illustrate the construction with the revolute joint.

\begin{figure}[htb]
  \centering
  \safeincludegraphics[width=0.6\linewidth]{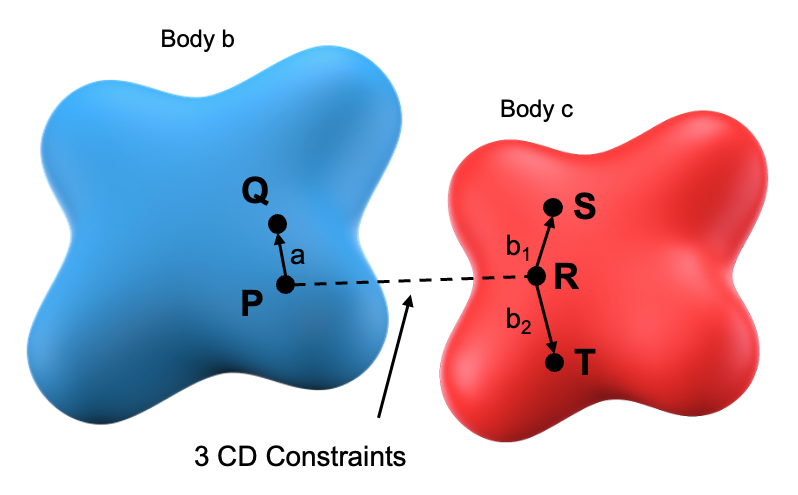}
  \caption{Geometry of the revolute joint construction used in this subsection. Points $\mathcal{P},\mathcal{Q}$ on body~$b$ define the body-attached axis direction $\mathbf{a}=\mathbf{r}_{\mathcal{Q}}-\mathbf{r}_{\mathcal{P}}$, while points $\mathcal{R},\mathcal{S},\mathcal{T}$ on body~$c$ define the off-axis directions $\mathbf{b}_1=\mathbf{r}_{\mathcal{S}}-\mathbf{r}_{\mathcal{R}}$ and $\mathbf{b}_2=\mathbf{r}_{\mathcal{T}}-\mathbf{r}_{\mathcal{R}}$. Three CD constraints enforce point coincidence at the hinge location ($\mathbf{r}_{\mathcal{P}}=\mathbf{r}_{\mathcal{R}}$), and two DP1 constraints prescribe the relative orientation through $\mathbf{a}^{\mathsf{T}}\mathbf{b}_1=f_1$ and $\mathbf{a}^{\mathsf{T}}\mathbf{b}_2=f_2$, leaving one rotational degree of freedom about the hinge axis in the regular, full-rank case}
  \label{fig:revolute_joint_geometry}
\end{figure}

\paragraph{Revolute-joint point set}
Five material points define the joint. On body~$b$: the attachment point~$\mathcal P$ (element~$E$) at the hinge location, and a nearby point~$\mathcal Q$ (element~$F$) such that the body-attached direction $\mathbf a=\mathbf r_{\mathcal Q}-\mathbf r_{\mathcal P}$ is aligned with the intended hinge axis in the reference configuration. This requires $\mathcal P$ and $\mathcal Q$ to be distinct, so that $\mathbf a\neq \mathbf 0$. On body~$c$: the attachment point~$\mathcal R$ (element~$G$) at the hinge location, and two off-axis points~$\mathcal S$ and~$\mathcal T$ (elements~$H$ and~$K$) defining directions $\mathbf b_1=\mathbf r_{\mathcal S}-\mathbf r_{\mathcal R}$ and $\mathbf b_2=\mathbf r_{\mathcal T}-\mathbf r_{\mathcal R}$. The directions $\mathbf b_1$ and $\mathbf b_2$ must be linearly independent and not both parallel to the hinge axis. Under these regularity conditions, the five scalar constraints introduced below are generically independent. The revolute-joint geometry is illustrated in Fig.~\ref{fig:revolute_joint_geometry}.

\paragraph{Constraint equations}
The revolute joint is defined by $m=5$ scalar constraints: three CD constraints for point coincidence and two DP1 constraints for relative orientation,
\begin{equation}
  \label{eq:rev_constraints}
  \mathbf c^{\mathrm{rev}}
  =
  \begin{bmatrix}
    \mathbf e_x^{\mathsf T}(\mathbf r_{\mathcal R}-\mathbf r_{\mathcal P}) \\[2pt]
    \mathbf e_y^{\mathsf T}(\mathbf r_{\mathcal R}-\mathbf r_{\mathcal P}) \\[2pt]
    \mathbf e_z^{\mathsf T}(\mathbf r_{\mathcal R}-\mathbf r_{\mathcal P}) \\[2pt]
    \mathbf a^{\mathsf T}\mathbf b_1 - f_1 \\[2pt]
    \mathbf a^{\mathsf T}\mathbf b_2 - f_2
  \end{bmatrix}
  = \mathbf 0 \in \mathbb{R}^5,
\end{equation}
where $f_j=\mathbf a_0^{\mathsf T}\mathbf b_{j,0}$ are constants precomputed from the reference configuration. Under the regularity conditions stated above, and provided the assembled joint Jacobian has local rank five, the three CD rows remove the three relative translations at the hinge, while the two DP1 rows remove two relative orientation components, leaving one rotational degree of freedom about~$\mathbf a$. When the off-axis directions are chosen perpendicular to the axis (i.e., $f_1=f_2=0$), the DP1 constraints reduce to $\mathbf a^{\mathsf T}\mathbf b_1=0$ and $\mathbf a^{\mathsf T}\mathbf b_2=0$, forcing $\mathbf a \parallel (\mathbf b_1\times\mathbf b_2)$.

\paragraph{Accumulated Jacobian blocks}
At the five constraint points, define the evaluation operators
\begin{gather}
  \boldsymbol\Sigma_E:=\mathbf s_E^b(\mathbf u_{\mathcal P})^{\mathsf T}\otimes \mathbf I_3,\qquad
  \boldsymbol\Sigma_F:=\mathbf s_F^b(\mathbf u_{\mathcal Q})^{\mathsf T}\otimes \mathbf I_3,\qquad
  \boldsymbol\Sigma_G:=\mathbf s_G^c(\mathbf u_{\mathcal R})^{\mathsf T}\otimes \mathbf I_3, \notag\\
  \boldsymbol\Sigma_H:=\mathbf s_H^c(\mathbf u_{\mathcal S})^{\mathsf T}\otimes \mathbf I_3,\qquad
  \boldsymbol\Sigma_K:=\mathbf s_K^c(\mathbf u_{\mathcal T})^{\mathsf T}\otimes \mathbf I_3.
  \label{eq:eval_op}
\end{gather}
Since $\mathcal P$ participates in both the CD and DP1 rows, and $\mathcal R$ does likewise, the corresponding hosting elements receive accumulated $5\times 3n_E$ and $5\times 3n_G$ Jacobian blocks:
\begin{equation}
  \label{eq:rev_accum_PR}
  \frac{\partial \mathbf c^{\mathrm{rev}}}{\partial \mathbf e_E^b}
  = -\begin{bmatrix}
      \mathbf e_x^{\mathsf T} \\[2pt]
      \mathbf e_y^{\mathsf T} \\[2pt]
      \mathbf e_z^{\mathsf T} \\[2pt]
      \mathbf b_1^{\mathsf T} \\[2pt]
      \mathbf b_2^{\mathsf T}
    \end{bmatrix}
    \boldsymbol\Sigma_E,
  \qquad
  \frac{\partial \mathbf c^{\mathrm{rev}}}{\partial \mathbf e_G^c}
  = \begin{bmatrix}
      +\mathbf e_x^{\mathsf T} \\[2pt]
      +\mathbf e_y^{\mathsf T} \\[2pt]
      +\mathbf e_z^{\mathsf T} \\[2pt]
      -\mathbf a^{\mathsf T} \\[2pt]
      -\mathbf a^{\mathsf T}
    \end{bmatrix}
    \boldsymbol\Sigma_G.
\end{equation}
The remaining elements contribute only through the DP1 rows:
\begin{equation}
  \label{eq:rev_accum_QST}
  \frac{\partial \mathbf c^{\mathrm{rev}}}{\partial \mathbf e_F^b}
  = \begin{bmatrix}
      \mathbf 0^{\mathsf T} \\[1pt]
      \mathbf 0^{\mathsf T} \\[1pt]
      \mathbf 0^{\mathsf T} \\[1pt]
      \mathbf b_1^{\mathsf T} \\[1pt]
      \mathbf b_2^{\mathsf T}
    \end{bmatrix}
    \boldsymbol\Sigma_F,\qquad
  \frac{\partial \mathbf c^{\mathrm{rev}}}{\partial \mathbf e_H^c}
  = \begin{bmatrix}
      \mathbf 0^{\mathsf T} \\[1pt]
      \mathbf 0^{\mathsf T} \\[1pt]
      \mathbf 0^{\mathsf T} \\[1pt]
      \mathbf a^{\mathsf T} \\[1pt]
      \mathbf 0^{\mathsf T}
    \end{bmatrix}
    \boldsymbol\Sigma_H,\qquad
  \frac{\partial \mathbf c^{\mathrm{rev}}}{\partial \mathbf e_K^c}
  = \begin{bmatrix}
      \mathbf 0^{\mathsf T} \\[1pt]
      \mathbf 0^{\mathsf T} \\[1pt]
      \mathbf 0^{\mathsf T} \\[1pt]
      \mathbf 0^{\mathsf T} \\[1pt]
      \mathbf a^{\mathsf T}
    \end{bmatrix}
    \boldsymbol\Sigma_K.
\end{equation}
These expressions make explicit how the primitive-level Jacobian rows combine into a composite-joint Jacobian. In particular, the algebraic coupling at the shared attachment points is already present at the element-block level; the global joint Jacobian is obtained by scattering these local blocks into the global row structure.

\paragraph{Constraint row normalization}
A mixed CD/DP1 constraint set introduces a scale mismatch. CD constraint values are $\mathcal O(\|\Delta\mathbf r\|)$, whereas DP1 values scale as $\mathcal O(\delta^2)$ when the direction vectors are defined by material fibers of length $\delta$, typically chosen as a small fraction of the element diameter. The corresponding Jacobian entries scale as $\mathcal O(1)$ for CD and $\mathcal O(\delta)$ for DP1. Without correction, the translational rows dominate both the residual and the Gauss--Newton term, and the angular part of the Newton step becomes poorly conditioned. We therefore apply constant row weights computed once from the reference configuration:
\begin{equation}
  \label{eq:row_weights}
  w_k = 1 \qquad (k=1,2,3;\;\text{CD}),\qquad
  w_{3+j}=\frac{1}{\sqrt{|\mathbf a_0|^2+|\mathbf b_{j,0}|^2}} \qquad (j=1,2;\;\text{DP1}).
\end{equation}
With $|\mathbf a_0|=|\mathbf b_{j,0}|=\delta$, one obtains $w_{3+j}\approx 1/\delta$. If $W_{\mathrm{rev}}=\mathrm{diag}(w_1,\ldots,w_5)$, the row-normalized revolute constraint vector and element Jacobian blocks are
\begin{equation}
  \label{eq:rev_normalized}
  \widehat{\mathbf c}^{\mathrm{rev}} = W_{\mathrm{rev}}\,\mathbf c^{\mathrm{rev}},\qquad
  \widehat{J}^{\mathrm{rev}}_X = W_{\mathrm{rev}}\,\frac{\partial \mathbf c^{\mathrm{rev}}}{\partial \mathbf e_X},\qquad X\in\{E,F,G,H,K\}.
\end{equation}
This normalization balances the translational and angular rows at the level of both residual and Jacobian. A more detailed scaling argument is given in Appendix~\ref{sec:joint_details}.

\paragraph{Residual insertion and consistent linearization}
The normalized blocks in \cref{eq:rev_normalized} enter the augmented-Lagrangian residual through an element-wise scatter. If $\widehat{\boldsymbol\eta}^{\mathrm{rev}}=\widehat{\boldsymbol\lambda}^{\mathrm{rev}}+\rho\,\widehat{\mathbf c}^{\mathrm{rev}}$, each hosting element receives
\begin{equation}
  \label{eq:rev_scatter}
  \mathbf g_X \;\mathrel{+}=\; h\,\bigl(\widehat{J}^{\mathrm{rev}}_X\bigr)^{\mathsf T}\widehat{\boldsymbol\eta}^{\mathrm{rev}},\qquad X\in\{E,F,G,H,K\}.
\end{equation}
Accordingly, the global transpose action $\mathbf C_q^{\mathsf T}\widehat{\boldsymbol\eta}$ is never formed as a dense matrix; it is assembled element-by-element in exactly the same way as internal and external force contributions.

The Newton system requires the derivative of the constraint contribution with respect to $\mathbf v$. With $\mathbf q=\mathbf q_n+h\,\mathbf v$ and with $\mathbf c$ and $\mathbf C_q$ denoting the row-normalized constraints and Jacobian used in the solve, one obtains
\begin{equation}
  \label{eq:newton_hessian}
  \frac{\partial\mathbf g_{\mathrm{con}}}{\partial\mathbf v}
  =
  \underbrace{h^2\rho\,\mathbf C_q^{\mathsf T}\mathbf C_q}_{\displaystyle\mathbf H_{\mathrm{GN}}}
  +
  \underbrace{h^2\sum_{i=1}^{m}\eta_i\,\nabla_q^2 c_i}_{\displaystyle\mathbf H_{\mathrm{curv}}},
\end{equation}
where $\eta_i=\lambda_i+\rho\,c_i$. Each CD constraint is linear in $\mathbf q$, so its Hessian vanishes. A DP1 constraint is quadratic in $\mathbf q$, and its Hessian is therefore constant, sparse, and in general indefinite. The explicit block form is given in Appendix~\ref{sec:joint_details}. The Gauss--Newton contribution $\mathbf H_{\mathrm{GN}}$ is positive semidefinite. The curvature correction $\mathbf H_{\mathrm{curv}}$ is the term required for a consistent Newton linearization, but because it is indefinite, the full Newton matrix
\begin{equation}
  \label{eq:full_newton}
  \mathbf H
  = \frac{1}{h}\mathbf M
  + h\,\mathbf K_t
  + h^2\rho\,\mathbf C_q^{\mathsf T}\mathbf C_q
  + h^2\textstyle\sum_{i=1}^{m}\eta_i\,\nabla_q^2 c_i
\end{equation}
is not guaranteed positive definite. Here $\mathbf K_t=\partial\mathbf f_{\mathrm{int}}/\partial\mathbf q$ is the tangent stiffness matrix. Retaining the curvature term yields the consistent linearization and quadratic convergence of Newton's method, but it generally requires a symmetric-indefinite factorization rather than Cholesky. Omitting $\mathbf H_{\mathrm{curv}}$ preserves the SPD structure at the cost of superlinear rather than quadratic convergence. These features are inherited by every joint that contains dot-product constraints.

\paragraph{Other standard joints}
The same regularity caveat applies to the remaining entries of \cref{tab:joint_decomp}: the stated DOF counts are local counts under full-row-rank configurations. A \emph{spherical} joint uses only the three CD constraints of the revolute construction, leaving three rotational degrees of freedom. A \emph{universal} (Hooke) joint adds one DP1 constraint $\mathbf a^{\mathsf T}\mathbf b_1=0$ to the spherical set, removing one off-axis rotation. A \emph{fixed} (weld) joint augments the revolute set with a third DP1 constraint, locking all relative rotation ($m=6$, zero DOF). A \emph{cylindrical} joint omits the CD block; two DP1 constraints enforce axis parallelism and two DP2 constraints enforce offset collinearity with the shared axis, leaving one axial rotation and one axial translation. A \emph{prismatic} joint adds one further DP1 constraint to the cylindrical set, locking axial rotation and retaining only one translational DOF along the shared axis.

\section{Stress Measures and Material Models Aspects}
\label{sec:mat_models_aspects}
\subsection{Stress Measures}

In the Total Lagrangian setting, all kinematic and constitutive quantities are referred to the reference configuration. The stress measure used later for internal-force and reference-surface-traction calculations is the first Piola--Kirchhoff stress $\mathbf{P}$, since it maps reference normals to current forces. The Cauchy stress $\boldsymbol{\sigma}$ acts on surfaces in the current configuration:
\begin{equation}
  \label{eq:cauchy-stress-force}
  d\vec{f} = \boldsymbol{\sigma}\,\vec{n}\,dA \, .
\end{equation}
The corresponding reference-area description is
\begin{equation}
  \label{eq:nanson-formula}
  \vec{n}\,dA = J\,\mathbf{F}^{-T}\vec{n}_0\,dA_0 \, ,
\end{equation}
which yields
\begin{equation}
  \label{eq:first-piola-kirchhoff-stress-force}
  d\vec{f} = \mathbf{P}\,\vec{n}_0\,dA_0 \, , \qquad \mathbf{P} = J\,\boldsymbol{\sigma}\,\mathbf{F}^{-T} \, .
\end{equation}
Accordingly, the traction per unit reference area is $\mathbf{t}_0=\mathbf{P}\mathbf{n}_0$.

For hyperelastic constitutive modeling it is also convenient to introduce the second Piola--Kirchhoff stress $\mathbf{S}$, defined by
\begin{equation}
  \label{eq:second-piola-kirchhoff-stress-definition}
  \mathbf{S} = J\,\mathbf{F}^{-1}\boldsymbol{\sigma}\,\mathbf{F}^{-T} \, ,
\end{equation}
so that
\begin{equation}
  \label{eq:stress-relations}
  J\,\boldsymbol{\sigma} = \mathbf{P}\,\mathbf{F}^{T} = \mathbf{F}\,\mathbf{S}\,\mathbf{F}^{T} \, , \qquad \mathbf{P}=\mathbf{F}\mathbf{S} \, .
\end{equation}
The stress-power identity associated with these three measures is
\begin{equation}
  \label{eq:stress-power-identity-inf-volume}
  J\,\boldsymbol{\sigma}:\sbelRateOfDeformation = \mathbf{P}:\dot{\mathbf{F}} = \mathbf{S}:\dot{\mathbf{E}} \, ,
\end{equation}
where $\sbelRateOfDeformation$ is the spatial rate-of-deformation tensor and $\mathbf{E}=\tfrac12(\mathbf{F}^T\mathbf{F}-\mathbf{I})$ is the Green--Lagrange strain. The corresponding virtual variations needed later are
\begin{equation}
  \delta \mathbf{F} = \nabla_0 \delta \mathbf{r} \, , \qquad \delta \mathbf{E} = \frac12\left(\delta \mathbf{F}^{T}\mathbf{F} + \mathbf{F}^{T}\delta \mathbf{F}\right) \, .
\end{equation}
Accordingly, the internal virtual work can be written in reference form as
\begin{equation}
  \label{eq:internal-virtual-work}
  \delta W_{\text{int}} = \int_{\sbelVolumeRef} \mathbf{P}:\delta \mathbf{F}\,d\sbelVolumeRef = \int_{\sbelVolumeRef} \mathbf{S}:\delta \mathbf{E}\,d\sbelVolumeRef \, .
\end{equation}
For the remainder of the paper, the constitutive interface is $\mathbf{P}(\mathbf{F})$ and, for consistent linearization, its material derivative $\partial \mathbf{P}/\partial \mathbf{F}$.

\subsection{Material Models}
\label{subsec:material_models}

A constitutive model specifies how stress depends on the local deformation state. In the present setting, the relevant output is the first Piola--Kirchhoff stress $\mathbf{P}$ and, later, its derivative with respect to $\mathbf{F}$. For hyperelastic materials, one may equivalently define the model through a strain energy density function per unit reference volume, either as $\sbelStrainEnergyDensityFunction(\mathbf{E})$ or as $\sbelStrainEnergyDensityFunction(\mathbf{F})$, with
\[
\mathbf{S} = \frac{\partial \sbelStrainEnergyDensityFunction}{\partial \mathbf{E}} \, , \qquad \mathbf{P} = \frac{\partial \sbelStrainEnergyDensityFunction}{\partial \mathbf{F}} \, .
\]
The models below are used as representative elastic and viscoelastic cases.

\subsubsection{The St.\ Venant--Kirchhoff (SVK) material model}
\label{sec:svk-material}

The SVK model is the finite-kinematics extension of linear elasticity obtained by replacing the infinitesimal strain with the Green--Lagrange strain while keeping the constitutive law linear in $\mathbf{E}$. Its strain energy density function is
\begin{equation}
  \Psi(\mathbf{E}) = \frac{\lambda}{2}\left(\operatorname{tr}\mathbf{E}\right)^2 + \mu\,\operatorname{tr}\left(\mathbf{E}^2\right) \, ,
\end{equation}
where $\lambda$ and $\mu$ are the Lam\'e constants. The associated second Piola--Kirchhoff stress is
\begin{equation}
  \mathbf{S} = \lambda\,\operatorname{tr}(\mathbf{E})\,\mathbf{I} + 2\mu\,\mathbf{E} \, ,
\end{equation}
and the first Piola--Kirchhoff stress is
\begin{equation}
  \mathbf{P} = \mathbf{F}\mathbf{S} = \lambda\,\operatorname{tr}(\mathbf{E})\,\mathbf{F} + 2\mu\,\mathbf{F}\mathbf{E} \, .
\end{equation}
Using $\mathbf{E}=\tfrac12(\mathbf{F}^T\mathbf{F}-\mathbf{I})$, one obtains
\begin{equation}
  \label{eq:first-piola-kirchhoff-stress-svk}
  \mathbf{P} = \lambda\left(\tfrac12\operatorname{tr}(\mathbf{F}^T\mathbf{F})-\tfrac32\right)\mathbf{F} + \mu\,\mathbf{F}\mathbf{F}^T\mathbf{F} - \mu\,\mathbf{F} \, .
\end{equation}
The model is simple and analytically convenient, but it is reliable only for small to moderate strains.

\subsubsection{The Mooney--Rivlin (M--R) material model}
\label{sec:mooney-rivlin-material}

The compressible Mooney--Rivlin model is a phenomenological hyperelastic law for nearly incompressible rubber-like materials. Let
\begin{align}
  \sbelRightCauchyGreenTensor &= \mathbf{F}^{T}\mathbf{F} \, , \\
  I_1 &= \operatorname{tr}(\sbelRightCauchyGreenTensor) \, , \\
  I_2 &= \frac12\left[\left(\operatorname{tr}\sbelRightCauchyGreenTensor\right)^2 - \operatorname{tr}\left(\sbelRightCauchyGreenTensor^2\right)\right] \, , \\
  J &= \det \mathbf{F} \, , \\
  \bar{I}_1 &= J^{-2/3} I_1 \, , \\
  \bar{I}_2 &= J^{-4/3} I_2 \, .
\end{align}
The strain energy density function is
\begin{equation}
  \Psi(\mathbf{F}) = \mu_{10}\left(\bar{I}_1-3\right) + \mu_{01}\left(\bar{I}_2-3\right) + \frac{k}{2}(J-1)^2 \, ,
\end{equation}
where $\mu_{10}$ and $\mu_{01}$ control the isochoric response and $k$ is the bulk modulus. The model reduces to neo-Hookean elasticity when $\mu_{01}=0$. Differentiation with respect to $\mathbf{F}$ yields
\begin{multline}
  \label{eq:first-piola-kirchhoff-stress-mooney-rivlin-final}
  \sbelFirstPiolaKirchhoffStress = 2\,\mu_{10}\,J^{-2/3}\!\left(\sbelDefGradient-\tfrac{1}{3} I_1\,\sbelDefGradient^{-T}\right) + 2\,\mu_{01}\,J^{-4/3}\!\left(I_1\sbelDefGradient-\sbelDefGradient\sbelRightCauchyGreenTensor-\tfrac{2}{3} I_2\,\sbelDefGradient^{-T}\right) \\
  \hfill + k\,(J-1)\,J\,\sbelDefGradient^{-T} \, . \hfill
\end{multline}

\subsubsection{The Kelvin--Voigt viscoelastic material}
\label{subsubsec:KV-viscoelastic-material}

To augment a hyperelastic model with rate dependence, we split the second Piola--Kirchhoff stress into elastic and viscous parts. Let $\sbelStrainEnergyDensityFunction(\sbelGreenLagrangeStrainTensor)$ denote a hyperelastic free-energy density. The elastic stresses are
\[
{\sbelSecondPiolaKirchhoffStress}^{\mathrm e} = \frac{\partial \sbelStrainEnergyDensityFunction}{\partial \sbelGreenLagrangeStrainTensor} \, , \qquad {\sbelFirstPiolaKirchhoffStress}^{\mathrm e} = \sbelDefGradient\,{\sbelSecondPiolaKirchhoffStress}^{\mathrm e} \, .
\]
For the viscous branch, introduce the dissipation potential
\[
\sbelDampingEnergyDensityFunction(\dot{\sbelGreenLagrangeStrainTensor}) = \mu_v\,\dot{\sbelGreenLagrangeStrainTensor}:\dot{\sbelGreenLagrangeStrainTensor} + \frac{\lambda_v}{2}\big(\operatorname{tr}\dot{\sbelGreenLagrangeStrainTensor}\big)^2 \, ,
\]
which yields the viscous stresses
\begin{subequations}
\label{eq:viscous2PKstress-subequations}
\begin{equation}
\label{eq:viscous2PKstress}
{\sbelSecondPiolaKirchhoffStress}^{\mathrm v} = \frac{\partial \sbelDampingEnergyDensityFunction}{\partial \dot{\sbelGreenLagrangeStrainTensor}} = 2\mu_v\,\dot{\sbelGreenLagrangeStrainTensor} + \lambda_v\,\operatorname{tr}(\dot{\sbelGreenLagrangeStrainTensor})\,\mathbf{I}_3 \, ,
\end{equation}
\begin{equation}
\label{eq:viscous1PKstress}
{\sbelFirstPiolaKirchhoffStress}^{\mathrm v} = \mathbf{F}\,{\sbelSecondPiolaKirchhoffStress}^{\mathrm v} \, .
\end{equation}
\end{subequations}
The total stresses are therefore
\begin{subequations}
\label{eq:total2PKstress-subequations}
\begin{equation}
{\sbelSecondPiolaKirchhoffStress} = {\sbelSecondPiolaKirchhoffStress}^{\mathrm e} + {\sbelSecondPiolaKirchhoffStress}^{\mathrm v} \, ,
\end{equation}
\begin{equation}
\label{eq:total1PKstress}
{\sbelFirstPiolaKirchhoffStress} = {\sbelFirstPiolaKirchhoffStress}^{\mathrm e} + {\sbelFirstPiolaKirchhoffStress}^{\mathrm v} = \mathbf{F}\,{\sbelSecondPiolaKirchhoffStress} \, .
\end{equation}
\end{subequations}

The definition of internal force does not change; only the constitutive input $\mathbf{P}$ becomes richer:
\begin{equation}
\label{eq:internalForceKelvinVoigtTotal}
\mathbf{f}_i^{\mathrm{int}} = \int_{\sbelVolumeRef} {\sbelFirstPiolaKirchhoffStress}\,{\sbelShapeFunElementGrad}_i\,d\sbelVolumeRef = \int_{\sbelVolumeRef} {\sbelFirstPiolaKirchhoffStress}^{\mathrm e}\,{\sbelShapeFunElementGrad}_i\,d\sbelVolumeRef + \int_{\sbelVolumeRef} {\sbelFirstPiolaKirchhoffStress}^{\mathrm v}\,{\sbelShapeFunElementGrad}_i\,d\sbelVolumeRef = \mathbf{f}_i^{\mathrm{int},e} + \mathbf{f}_i^{\mathrm{int},v} \, .
\end{equation}
At each quadrature point, the procedure is therefore: compute $\mathbf{E}$ and $\dot{\mathbf{E}}$, evaluate the elastic stress from the chosen hyperelastic model, evaluate the viscous stress from \cref{eq:viscous2PKstress}, and form the total first Piola--Kirchhoff stress through \cref{eq:total1PKstress}.

\paragraph{Thermodynamic consistency}
The Kelvin--Voigt branch is thermodynamically admissible provided the viscous dissipation is nonnegative:
\[
\mathcal{D} := {\sbelSecondPiolaKirchhoffStress}^{\mathrm v}:\dot{\sbelGreenLagrangeStrainTensor} = 2\mu_v\,\dot{\sbelGreenLagrangeStrainTensor}:\dot{\sbelGreenLagrangeStrainTensor} + \lambda_v\big(\operatorname{tr}\dot{\sbelGreenLagrangeStrainTensor}\big)^2 \, .
\]
Hence $\mathcal{D}\ge 0$ whenever $\mu_v\ge 0$ and $\lambda_v\ge 0$. Under this condition, the elastic part stores energy through $\sbelStrainEnergyDensityFunction$, the viscous part dissipates energy through $\sbelDampingEnergyDensityFunction$, and the additive stress split in \cref{eq:total1PKstress} is consistent with the Clausius--Duhem inequality.

\section{Dynamics Aspects}
\label{sec:dynamics_aspects}
This section derives the virtual-work contributions associated with inertia, mass-distributed force fields, internal forces, concentrated point loads, and prescribed surface tractions in the TL-FEA setting. The goal is to establish a consistent notation for the force terms that enter the equations of motion. With the sign convention adopted here, the virtual-work balance is written as
\begin{equation}
\label{eq:virtual_work_decomposition}
\delta W
=
\delta W_{\mathrm{inertia}}
+
\delta W_{\mathrm{applied}}
+
\delta W_{\mathrm{force\text{-}field}}
+
\delta W_{\mathrm{internal}}
=0 \; .
\end{equation}
Each contribution is expressed below as a linear form in the virtual nodal variations, which identifies the corresponding nodal force vectors after standard finite-element assembly.

\subsection{Inertia Force Contribution in TL-FEA Formulation}
\label{sec:inertia_contribution}

The inertia contribution in Eq.~\eqref{eq:virtual_work_decomposition} is
\begin{equation}
\label{eq:virtual_work_inertia}
\delta W_{\mathrm{inertia}}
= -\int_{V_r} \rho_r(\mathbf{u})\,\delta\mathbf{r}(\mathbf{u};t)^{T}\ddot{\mathbf{r}}(\mathbf{u};t)\,\mathrm{d}V_r \; ,
\end{equation}
where $\rho_r(\mathbf{u})$ is the reference density and $\mathbf{u}=[u,v,w]^T$ are the material coordinates. Using the TL interpolation
\[
\mathbf{r}(\mathbf{u};t)=\sum_{i=1}^{n}\mathbf{e}_i(t)\,s_i(\mathbf{u}),\qquad
\delta\mathbf{r}(\mathbf{u};t)=\sum_{i=1}^{n}\delta\mathbf{e}_i\,s_i(\mathbf{u}),\qquad
\ddot{\mathbf{r}}(\mathbf{u};t)=\sum_{j=1}^{n}\ddot{\mathbf{e}}_j\,s_j(\mathbf{u}),
\]
one obtains
\begin{equation}
\label{eq:virtual_work_inertia_discrete}
\delta W_{\mathrm{inertia}}
= -\sum_{i=1}^{n}\delta\mathbf{e}_i^{T}\sum_{j=1}^{n}m_{ij}\,\ddot{\mathbf{e}}_j \; ,
\qquad
m_{ij}
=
\int_{V_r}\rho_r(\mathbf{u})\,s_i(\mathbf{u})\,s_j(\mathbf{u})\,\mathrm{d}V_r \; .
\end{equation}
The coefficients $m_{ij}=m_{ji}$ depend only on reference quantities and are therefore constant for a given element.

Introducing the element unknown vector
\[
\mathbf{e}=[\mathbf{e}_1^{T},\ldots,\mathbf{e}_{n}^{T}]^{T}\in\mathbb{R}^{3n} \; ,
\]
Eq.~\eqref{eq:virtual_work_inertia_discrete} can be written as
\begin{equation}
\label{eq:element_mass_matrix}
\delta W_{\mathrm{inertia}}
=
-\delta\mathbf{e}^{T}\mathbf{M}_e\,\ddot{\mathbf{e}},
\qquad
\mathbf{M}_e=\bigl[m_{ij}\mathbf{I}_3\bigr]_{i,j=1}^{n} \; .
\end{equation}
After assembly over all elements, this yields the global inertial virtual work
\[
\delta W_{\mathrm{inertia}}=-\delta\mathbf{q}^{T}\mathbf{M}\,\ddot{\mathbf{q}},
\]
where $\mathbf{M}$ is the global consistent mass matrix.

\subsection{Mass-Distributed Force Field Contribution in TL-FEA Formulation}
\label{sec:mass_distributed_contribution}

Let $\mathbf{b}(P,t)\in\mathbb{R}^{3}$ denote a force per unit mass, evaluated at the current spatial position $P=\mathbf{r}(\mathbf{u};t)$ of the material point with reference coordinates $\mathbf{u}$. The corresponding virtual work over the reference volume is
\begin{equation}
\label{eq:virtual_work_force_field}
\delta W_{\mathrm{force\text{-}field}}
=
\int_{V_r}\rho_r(\mathbf{u})\,\delta\mathbf{r}(\mathbf{u};t)^{T}\mathbf{b}(\mathbf{r}(\mathbf{u};t),t)\,\mathrm{d}V_r \; .
\end{equation}
Using
\[
\delta\mathbf{r}(\mathbf{u};t)=\sum_{i=1}^{n}\delta\mathbf{e}_i\,s_i(\mathbf{u}),
\]
Eq.~\eqref{eq:virtual_work_force_field} becomes
\begin{equation}
\label{eq:virtual_work_force_field_discrete}
\delta W_{\mathrm{force\text{-}field}}
=
\sum_{i=1}^{n}\delta\mathbf{e}_i^{T}\mathbf{f}^{\mathrm{ff}}_i,
\qquad
\mathbf{f}^{\mathrm{ff}}_i
\coloneqq
\int_{V_r}\rho_r(\mathbf{u})\,s_i(\mathbf{u})\,\mathbf{b}(\mathbf{r}(\mathbf{u};t),t)\,\mathrm{d}V_r
\in\mathbb{R}^{3} \; .
\end{equation}
After element assembly, these contributions define the global force-field vector.

If $\rho_r$ is constant and the field is uniform, $\mathbf{b}(\mathbf{r}(\mathbf{u};t),t)\equiv\mathbf{b}_0$, then
\begin{equation}
\label{eq:virtual_work_force_field_uniform}
\mathbf{f}^{\mathrm{ff}}_i
=
\rho_r\,v_i\,\mathbf{b}_0,
\qquad
v_i
\coloneqq
\int_{V_r}s_i(\mathbf{u})\,\mathrm{d}V_r \; .
\end{equation}
Although the integration is carried out over $V_r$, the field is evaluated at the current spatial position $P=\mathbf{r}(\mathbf{u};t)$; consequently, $\mathbf{b}$ may vary in time as the body moves through space.

\subsection{Internal Force Contribution in TL-FEA Formulation}
\label{sec:internal_force_contribution}

The internal virtual work follows from the constitutive relation introduced in Section~\ref{sec:mat_models_aspects}:
\begin{equation}
\delta W_{\mathrm{internal}}
=
-\int_{V_r}\mathbf{P}(\mathbf{F}):\delta\mathbf{F}\,\mathrm{d}V_r \; .
\end{equation}
Using the representation
\[
\mathbf{F}(\mathbf{u};t)=\sum_{i=1}^{n}\mathbf{e}_i(t)\,\mathbf{h}_i(\mathbf{u})^{T},
\qquad
\delta\mathbf{F}(\mathbf{u};t)=\sum_{i=1}^{n}\delta\mathbf{e}_i\,\mathbf{h}_i(\mathbf{u})^{T},
\]
together with the identity $\mathbf{A}:(\mathbf{x}\mathbf{y}^{T})=\mathbf{x}^{T}\mathbf{A}\mathbf{y}$, one obtains
\[
\delta W_{\mathrm{internal}}
=
-\sum_{i=1}^{n}\delta\mathbf{e}_i^{T}
\int_{V_r}\mathbf{P}(\mathbf{F})\,\mathbf{h}_i(\mathbf{u})\,\mathrm{d}V_r \; .
\]
This identifies the internal force associated with nodal unknown $\mathbf{e}_i$ as
\begin{equation}
\label{eq:internal_force_universal}
\mathbf{f}_i^{\mathrm{int}}
=
\int_{V_r}\mathbf{P}(\mathbf{F})\,\mathbf{h}_i(\mathbf{u})\,\mathrm{d}V_r
\in\mathbb{R}^{3} \; ,
\end{equation}
where $\mathbf{h}_i=\nabla_{\mathbf{u}}s_i\in\mathbb{R}^{3}$ is the reference gradient of the $i$th shape function defined in Eq.~\eqref{eq:shape-func-def-grad-comp}.

Eq.~\eqref{eq:internal_force_universal} is the constitutive-discretization interface of the formulation. The element topology enters only through the gradients $\mathbf{h}_i$, while the constitutive response enters only through the first Piola--Kirchhoff stress $\mathbf{P}(\mathbf{F})$. For Kelvin--Voigt materials, $\mathbf{P}$ splits additively into elastic and viscous parts, which induces the same additive split in $\mathbf{f}_i^{\mathrm{int}}$. If a consistent tangent is needed, the derivative $\partial \mathbf{f}_i^{\mathrm{int}}/\partial \mathbf{e}_j$ depends on the material derivative $\partial \mathbf{P}/\partial \mathbf{F}$. Explicit expressions for $\mathbf{P}$ and $\partial\mathbf{P}/\partial\mathbf{F}$ for the SVK, Mooney--Rivlin, and Kelvin--Voigt models are provided in Appendix~\ref{sec:select_material_models}.

\subsection{Pointwise External Force Loads}
\label{sec:pointwise_external_loads}

Pointwise external loads act at material points and are distributed to the element nodal unknowns through the shape functions. Let $\mathbf{f}_{P}(t)\in\mathbb{R}^{3}$ denote a force applied at the material point $P$ of reference coordinates $\mathbf{u}_{P}=[u,v,w]^T$. Its virtual work contribution is
\begin{equation}
\label{eq:virtual_work_point_force}
\delta W_{\mathrm{applied}}^{P}
=
\delta\mathbf{r}(\mathbf{u}_{P};t)^{T}\mathbf{f}_{P}(t) \; .
\end{equation}
Using
\[
\delta\mathbf{r}(\mathbf{u}_{P};t)=\sum_{i=1}^{n}\delta\mathbf{e}_i\,s_i(\mathbf{u}_{P}),
\]
Eq.~\eqref{eq:virtual_work_point_force} becomes
\begin{equation}
\label{eq:virtual_work_point_force_discrete}
\delta W_{\mathrm{applied}}^{P}
=
\sum_{i=1}^{n}\delta\mathbf{e}_i^{T}\mathbf{f}_{iP},
\qquad
\mathbf{f}_{iP}
\coloneqq
s_i(\mathbf{u}_{P})\,\mathbf{f}_{P}(t)
\in\mathbb{R}^{3} \; .
\end{equation}
Therefore, a concentrated load is distributed to the nodal unknowns according to the shape functions evaluated at the application point. After assembly over all pointwise loads and all elements, these contributions define the global applied-load vector associated with concentrated forces.

\subsection{Surface Traction Loads}
\label{sec:surface_traction_loads}

Surface tractions represent applied forces distributed over loaded boundary surfaces. Let $\Gamma_{t,0}\subset\partial V_r$ denote the loaded portion of the boundary in the reference configuration, and let $\bar{\mathbf{t}}_0(\mathbf{u},t)\in\mathbb{R}^{3}$ denote the prescribed \emph{nominal traction}, i.e., the force per unit reference area. Its virtual work is
\begin{equation}
\label{eq:virtual_work_traction}
\delta W_{\mathrm{traction}}
=
\int_{\Gamma_{t,0}}
\delta\mathbf{r}(\mathbf{u};t)^{T}\,\bar{\mathbf{t}}_0(\mathbf{u},t)\,\mathrm{d}A_0 \; .
\end{equation}

If the load is prescribed instead as a traction $\bar{\mathbf{t}}$ per unit current area on the current surface, the corresponding nominal traction is obtained by pull-back to the reference surface. Since
\[
\mathrm{d}A = J\,\|\mathbf{F}^{-T}\mathbf{n}_0\|\,\mathrm{d}A_0,
\]
one has
\begin{equation}
\label{eq:nominal_traction}
\bar{\mathbf{t}}_0
=
J\,\|\mathbf{F}^{-T}\mathbf{n}_0\|\,\bar{\mathbf{t}} \; .
\end{equation}
In the special case where the applied traction is induced by a Cauchy stress, $\bar{\mathbf{t}}=\boldsymbol{\sigma}\mathbf{n}$, this reduces to $\bar{\mathbf{t}}_0=\mathbf{P}\mathbf{n}_0$, consistent with Eq.~\eqref{eq:first-piola-kirchhoff-stress-force}.

Using
\[
\delta\mathbf{r}(\mathbf{u};t)=\sum_{i=1}^{n}\delta\mathbf{e}_i\,s_i(\mathbf{u}),
\]
Eq.~\eqref{eq:virtual_work_traction} becomes
\begin{equation}
\label{eq:virtual_work_traction_discrete}
\delta W_{\mathrm{traction}}
=
\sum_{i=1}^{n}\delta\mathbf{e}_i^{T}\,\mathbf{f}_i^{\mathrm{tr}},
\qquad
\mathbf{f}_i^{\mathrm{tr}}
\coloneqq
\int_{\Gamma_{t,0}}s_i(\mathbf{u})\,\bar{\mathbf{t}}_0(\mathbf{u},t)\,\mathrm{d}A_0
\in\mathbb{R}^{3} \; .
\end{equation}
These surface integrals are evaluated by boundary quadrature over the loaded element faces, following the isoparametric procedure described in Section~\ref{sec:approximation_of_integrals}. After assembly over all loaded faces and all elements, they define the global applied-load vector associated with surface tractions.

\section{Frictional Contact Aspects}
\label{sec:frictional_contact_aspects}
In the proposed TL-FEA formulation, we handle friction and contact using a penalty-based contact law with damping and history-dependent tangential friction. Note that the model is piecewise (compressive-only normal response via $\max(0,\cdot)$ and stick--slip branching) and is therefore nonsmooth at regime transitions; if a differentiable operator is required, one may replace $\max(0,x)$ with $\tfrac{1}{2}(x+\sqrt{x^2+\varepsilon^2})$ and adopt a regularized Coulomb law. At each active contact, the contact geometry and kinematic quantities $\mathbf{x}_c$, $\delta$, $\mathbf{n}$, and $A_{\mathrm{patch}}$ are supplied by the contact-detection layer, while the constitutive law described below consumes these quantities together with the local relative velocity to evaluate the contact force and assemble it into the external force vector.

We adopt a damped Hertz-type model for the normal response and a Mindlin-inspired tangential spring-dashpot law with history and Coulomb capping for the tangential response. The latter is intended as a practical friction model for the present implementation, not as the full Mindlin--Deresiewicz theory with all of its path-dependence subtleties and normal-load coupling.

\subsection{Contact Kinematics and Activation}
Consider two bodies $\sbelBodySub{A}$ and $\sbelBodySub{B}$ in contact at point $\mathbf{x}_c$. Let $\mathbf{n}$ be the unit normal pointing from $\sbelBodySub{B}$ to $\sbelBodySub{A}$, and let $\delta$ denote the penetration depth (contact active when $\delta>0$). Define the relative velocity at the contact point as $\mathbf{v}_{\mathrm{rel}}=\mathbf{v}_{\sbelBodySub{A}}(\mathbf{x}_c)-\mathbf{v}_{\sbelBodySub{B}}(\mathbf{x}_c)$, with normal and tangential components
\begin{equation}
\label{eq:contact_vn_vt}
v_n=\mathbf{v}_{\mathrm{rel}}\cdot \mathbf{n},
\qquad
\mathbf{v}_t=\mathbf{v}_{\mathrm{rel}}-v_n\mathbf{n}.
\end{equation}
The contact force is applied in action--reaction form: $\mathbf{F}_{c,\sbelBodySub{A}}=\mathbf{F}_n+\mathbf{F}_t$ and $\mathbf{F}_{c,\sbelBodySub{B}}=-\mathbf{F}_{c,\sbelBodySub{A}}$. When $\delta\le 0$, we set $\mathbf{F}_n=\mathbf{F}_t=\mathbf{0}$ and reset the tangential history variable (Section~\ref{sec:mindlin_friction}).

\subsection{Damped Hertz-type normal law}
We use an effective elastic modulus~\cite{johnson1987contact}
\begin{equation}
\label{eq:contact_Eeff}
\frac{1}{E_{\mathrm{eff}}}=\frac{1-\nu_{\sbelBodySub{A}}^{2}}{E_{\sbelBodySub{A}}}+\frac{1-\nu_{\sbelBodySub{B}}^{2}}{E_{\sbelBodySub{B}}}.
\end{equation}
To accommodate both sphere-like and patch-based contacts, we introduce an effective contact radius $a$. For smooth spheres, $a=\sqrt{R_{\mathrm{eff}}\delta}$ with $R_{\mathrm{eff}}^{-1}=R_{\sbelBodySub{A}}^{-1}+R_{\sbelBodySub{B}}^{-1}$; for a triangle/patch contact of area $A_{\mathrm{patch}}$, we use the equivalent disk radius $a=\sqrt{A_{\mathrm{patch}}/\pi}$. The patch case is therefore an area-based extension used for general contact geometries and should be interpreted as a local penalty stiffness model rather than as a strict Hertz derivation. With the stiffness scale $S_n=2E_{\mathrm{eff}}a$ and coefficient $k_n=\tfrac{2}{3}S_n=\tfrac{4}{3}E_{\mathrm{eff}}a$, the elastic contribution is $k_n\delta$, which recovers the classical Hertz scaling $F_n\propto \delta^{3/2}$ when $a=\sqrt{R_{\mathrm{eff}}\delta}$.

To model dissipation, we use a restitution-based damping parameter~\cite{Brilliantov_1998,tsuji1992lagrangian}
\begin{equation}
\label{eq:contact_beta}
\beta=\frac{\ln(e)}{\sqrt{\ln(e)^2+\pi^2}}, \qquad e\in(0,1],
\end{equation}
and define the (nonnegative) normal damping coefficient
\begin{equation}
\label{eq:contact_gamman}
\gamma_n=-2\sqrt{\frac{5}{6}}\,\beta\,\sqrt{S_n\,m_{\mathrm{eff}}}.
\end{equation}
Here $m_{\mathrm{eff}}$ is taken in the present implementation as a body-level reduced-mass approximation rather than as a contact-direction effective mass extracted from the FE mass matrix; for scalar masses it reduces to $m_{\mathrm{eff}}=(m_{\sbelBodySub{A}}^{-1}+m_{\sbelBodySub{B}}^{-1})^{-1}$. The normal contact force is
\begin{equation}
\label{eq:contact_normal_force}
\mathbf{F}_n = F_n\,\mathbf{n},
\qquad
F_n=\max\!\left(0,\,k_n\delta-\gamma_n v_n\right),
\end{equation}
which enforces a compressive-only response and yields a repulsive damping contribution on approach ($v_n<0$).

\subsection{Mindlin history-dependent tangential friction}
\label{sec:mindlin_friction}
We model tangential friction using a Mindlin-inspired tangential spring-dashpot law with a history (stiction) variable~\cite{mindlinCompliance1949,mindlin53,jonJCND2015}. The model retains the key implementation ingredients needed here---incremental tangential stiffness, history dependence, and Coulomb capping---but is not intended as a full Mindlin--Deresiewicz contact theory. Let $\boldsymbol{\delta}_t$ denote the stored tangential spring displacement for each contact. At each step, we update and re-project it onto the current tangent plane,
\begin{equation}
\label{eq:contact_deltat_update}
\boldsymbol{\delta}_t \leftarrow \left(\mathbf{I}-\mathbf{n}\mathbf{n}^T\right)\left(\boldsymbol{\delta}_t+\Delta t\,\mathbf{v}_t\right).
\end{equation}
This projection is the adopted history transport rule in the present implementation. We define the effective shear modulus as $G_i=E_i/(2(1+\nu_i))$ and
\begin{equation}
\label{eq:contact_Geff}
\frac{1}{G_{\mathrm{eff}}}=\frac{2-\nu_{\sbelBodySub{A}}}{G_{\sbelBodySub{A}}}+\frac{2-\nu_{\sbelBodySub{B}}}{G_{\sbelBodySub{B}}}.
\end{equation}
The tangential stiffness and damping are
\begin{equation}
\label{eq:contact_kt_gammat}
k_t=8G_{\mathrm{eff}}a,
\qquad
\gamma_t=-2\sqrt{\frac{5}{6}}\,\beta\,\sqrt{m_{\mathrm{eff}}k_t},
\end{equation}
and the trial tangential force is
\begin{equation}
\label{eq:contact_tangential_trial}
\mathbf{F}_t^{\mathrm{trial}}=-k_t\boldsymbol{\delta}_t-\gamma_t\mathbf{v}_t.
\end{equation}
With a Coulomb limit $F_{\max}=\mu \|\mathbf{F}_n\|$, we set
\begin{equation}
\label{eq:contact_stick_slip}
\mathbf{F}_t=
\begin{cases}
\mathbf{F}_t^{\mathrm{trial}}, & \|\mathbf{F}_t^{\mathrm{trial}}\|\le F_{\max}\quad (\text{stick}),\\[0.2em]
\dfrac{F_{\max}}{\|\mathbf{F}_t^{\mathrm{trial}}\|}\,\mathbf{F}_t^{\mathrm{trial}}, & \|\mathbf{F}_t^{\mathrm{trial}}\|> F_{\max}\quad (\text{slip}).
\end{cases}
\end{equation}
In the slip regime, we rewind the history variable to keep $\mathbf{F}_t=-k_t\boldsymbol{\delta}_t-\gamma_t\mathbf{v}_t$ consistent with the capped force:
\begin{equation}
\label{eq:contact_deltat_rewind}
\boldsymbol{\delta}_t \leftarrow -\frac{\mathbf{F}_t+\gamma_t\mathbf{v}_t}{k_t}.
\end{equation}
This update reduces chatter by ensuring that the next step starts from the Coulomb boundary rather than from an over-extended tangential spring.

Finally, once $\mathbf{F}_n$ and $\mathbf{F}_t$ are evaluated at the contact point, they are treated as instantaneous external forces applied to the corresponding element(s) in action--reaction form. The resulting nodal load distribution follows the same procedure as for pointwise external force loads in Section~\ref{sec:pointwise_external_loads}.

\section{Approximation of Integrals}
\label{sec:approximation_of_integrals}
For definiteness, this section is written for three-dimensional volume elements, for which the parent coordinates are $\boldsymbol{\xi}=(\xi,\eta,\zeta)$. The same chain-rule construction applies to lower-dimensional structural elements, with the obvious dimensional changes in the parent-to-reference Jacobian.

For an isoparametric element, the mapping from parent coordinates $\boldsymbol{\xi}$ to reference coordinates $\mathbf{u}=(u,v,w)$ is
\begin{subequations}
\label{eq:isoparametricMapping}
\begin{equation}
  \mathbf{u}(\boldsymbol{\xi}) = \mathbf{U}\,{\sbelShapeFunArrayParent(\boldsymbol{\xi})},
\end{equation}
where $\sbelShapeFunArrayParent(\boldsymbol{\xi}) \in \mathbb{R}^{\sbelNUcount}$ is the vector of shape functions defined over the parent element, and $\mathbf{U} \in \mathbb{R}^{3\times \sbelNUcount}$ is the matrix of nodal coordinates of the reference element,
\begin{equation}
    \mathbf{U} \coloneqq 
    \begin{bmatrix}
     \mathbf{u}_1 & \mathbf{u}_2 & \ldots & \mathbf{u}_{\sbelNUcount} 
   \end{bmatrix} \in \mathbb{R}^{3\times {\sbelNUcount}} \; .
\end{equation}
The matrix $\mathbf{U}$ is constant: it is assembled once, column by column, with column $i$ storing the location of node $i$ of the reference element in the reference coordinate system.
\end{subequations}

In the isoparametric setting, the same parent shape functions are used both to describe the geometry of the reference element and to interpolate the field variable. In the present case, the field variable is the current position of a point. By slight abuse of notation, we write $\mathbf r(\boldsymbol{\xi};t)$ for the composed map $\mathbf r(\mathbf u(\boldsymbol{\xi});t)$ from the parent element to the current configuration:
\begin{equation}
    \mathbf{r}(\boldsymbol{\xi};t) 
    = 
    {\sbelNUmatrix(t) \: {\sbelShapeFunArrayParent(\boldsymbol{\xi})}} \; ,
\end{equation}
where, as before, $\sbelNUmatrix(t) \in \mathbb{R}^{3 \times \sbelNUcount}$ is the matrix of nodal unknowns at time $t$.

The deformation measures introduced earlier, such as $\sbelDefGradient$, $\sbelRightCauchyGreenTensor$, and $\sbelGreenLagrangeStrainTensor$, depend on derivatives with respect to the reference coordinates $\mathbf u$. Since the interpolation is evaluated in parent coordinates, these derivatives are obtained through the chain rule.

\begin{figure}[t]
\centering
\safeincludegraphics[width=0.4\textwidth]{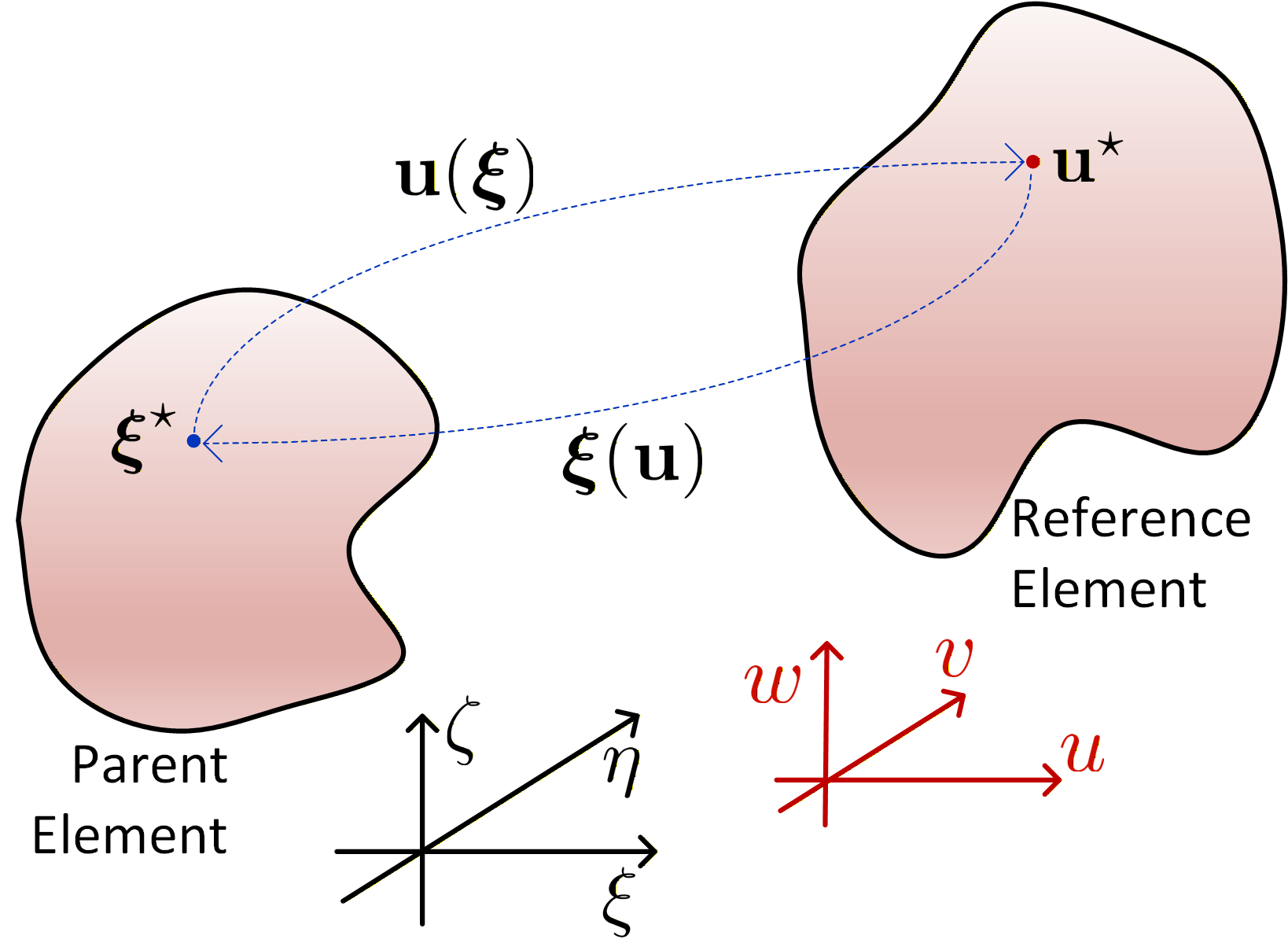}
\caption{The one-to-one mapping between the parent and reference configurations}
\label{fig:parent-to-reference}
\end{figure}

As shown in \cref{fig:parent-to-reference}, we assume that the parent-to-reference map is one-to-one and sufficiently regular. More precisely, for every evaluation point $\boldsymbol{\xi}^\star$ in the parent element, the Jacobian
\[
\frac{\partial \mathbf{u}}{\partial \boldsymbol{\xi}}(\boldsymbol{\xi}^\star)
\]
is assumed invertible, and for a valid element map one requires
\[
J_p(\boldsymbol{\xi}) \coloneqq \det\!\left(\frac{\partial \mathbf{u}}{\partial \boldsymbol{\xi}}(\boldsymbol{\xi})\right) > 0
\]
throughout the parent domain. Under this assumption, each point $\boldsymbol{\xi}^\star$ in the parent element corresponds to a unique point $\mathbf{u}^\star=\mathbf{u}(\boldsymbol{\xi}^\star)$ in the reference element, and vice versa. Hence, the reference-element shape functions satisfy
\[
\sbelShapeFun_i(\mathbf{u}^\star) 
= 
\sbelShapeFun_i(\mathbf{u}(\boldsymbol{\xi}^\star)) 
\coloneqq 
\sbelShapeFunParent_i(\boldsymbol{\xi}^\star) \; .
\]

Using the chain rule,
\[
\frac{\partial \sbelShapeFunParent_i}{\partial \boldsymbol{\xi}}(\boldsymbol{\xi}^\star) 
= 
\left[\frac{\partial \sbelShapeFun_i}{\partial \mathbf{u}}(\mathbf{u}^\star)\right]
\left[\frac{\partial \mathbf{u}}{\partial \boldsymbol{\xi}}(\boldsymbol{\xi}^\star)\right] \; ,
\]
and therefore
\begin{equation}
\frac{\partial \sbelShapeFun_i}{\partial \mathbf{u}}(\mathbf{u}^\star) 
= 
\frac{\partial \sbelShapeFunParent_i}{\partial \boldsymbol{\xi}}(\boldsymbol{\xi}^\star) \left[\frac{\partial \mathbf{u}}{\partial \boldsymbol{\xi}}(\boldsymbol{\xi}^\star)\right]^{-1} 
\qquad \Rightarrow \qquad
\sbelShapeFunElementGrad^T_i(\mathbf{u}^\star)
=
\frac{\partial \sbelShapeFunParent_i}{\partial \boldsymbol{\xi}}(\boldsymbol{\xi}^\star)
\left[\mathbf{U} \; \sbelShapeFunJacParent(\boldsymbol{\xi}^\star)\right]^{-1}
\; .
\end{equation}
Here $\partial \sbelShapeFun_i/\partial \mathbf u$ is written as a row Jacobian, while $\sbelShapeFunElementGrad_i(\mathbf u^\star)=\nabla_{\mathbf u}\sbelShapeFun_i(\mathbf u^\star)$ denotes the corresponding column gradient introduced in Section~\ref{sec:kinematic_aspects}.

Applying the chain rule to the composed map $\mathbf r(\mathbf u(\boldsymbol{\xi});t)$ gives
\[
\frac{\partial \mathbf{r}}{\partial \boldsymbol{\xi}}(\boldsymbol{\xi}^\star) = \frac{\partial \mathbf{r}}{\partial \mathbf{u}}(\mathbf{u}^\star) \; \frac{\partial \mathbf{u}}{\partial \boldsymbol{\xi}}(\boldsymbol{\xi}^\star) \; ,
\]
and therefore the deformation gradient at any point $\mathbf u^\star$ in the reference element is
\begin{equation}
\begin{split}
    \sbelDefGradient(\mathbf{u}^\star)
    &= \frac{\partial \mathbf{r}}{\partial \mathbf{u}}(\mathbf{u}^\star)
    = \frac{\partial \mathbf{r}}{\partial \boldsymbol{\xi}}(\boldsymbol{\xi}^\star) \; \left(\frac{\partial \mathbf{u}}{\partial \boldsymbol{\xi}} (\boldsymbol{\xi}^\star)\right)^{-1} \\
    &= \sbelNUmatrix(t) \; \sbelShapeFunJacParent(\boldsymbol{\xi}^\star) \; \left[\mathbf{U} \; \sbelShapeFunJacParent(\boldsymbol{\xi}^\star)\right]^{-1}
    \coloneqq \sbelNUmatrix(t) \; \sbelShapeFunJac(\mathbf{u}^\star) \; .
\end{split}
\end{equation}

The quantities needed later for constitutive evaluation and quadrature are:
\begin{subequations}
\label{eq:importantQuantitiesParent2ReferenceJump}
\begin{itemize}
    \item The Jacobian of the shape-function array with respect to the reference coordinates, evaluated at $\mathbf u^\star$:
    \begin{equation}
        \sbelShapeFunJac(\mathbf{u}^\star) = \sbelShapeFunJacParent(\boldsymbol{\xi}^\star) \; \left[\mathbf{U} \; \sbelShapeFunJacParent(\boldsymbol{\xi}^\star)\right]^{-1} \; .
    \end{equation}
    \item The deformation gradient at $\mathbf u^\star$:
    \begin{equation}
        \sbelDefGradient(\mathbf{u}^\star) = \sbelNUmatrix(t) \; \sbelShapeFunJac(\mathbf{u}^\star) \; .
    \end{equation}
    \item The transpose of the $i$th row of $\sbelShapeFunJac(\mathbf u^\star)$, which appears in the internal-force expression associated with nodal unknown $\sbelNUSub{i}$:
    \begin{equation}
      \label{eq:sbelShapeFunElementGrad-usingParentInformation}
      \sbelShapeFunElementGrad^T_i(\mathbf{u}^\star)
      =
      \frac{\partial \sbelShapeFunParent_i}{\partial \boldsymbol{\xi}}(\boldsymbol{\xi}^\star)
      \left[\mathbf{U} \; \sbelShapeFunJacParent(\boldsymbol{\xi}^\star)\right]^{-1} \; .
    \end{equation}
    \item The scaling factor for integration over the reference element:
    \begin{equation}
        J_p(\boldsymbol{\xi}^\star) \coloneqq \det\!\left(\frac{\partial \mathbf{u}}{\partial \boldsymbol{\xi}} (\boldsymbol{\xi}^\star)\right) 
        =
        \det\!\left[\mathbf{U} \; \sbelShapeFunJacParent(\boldsymbol{\xi}^\star)\right] \; .
    \end{equation}
\end{itemize}
\end{subequations}

An integral over the reference element is evaluated by a change of variables followed by quadrature:
\begin{subequations}
\label{eq:approximation-of-integral-over-reference-element}
\begin{equation}
    \int_{\sbelVolumeRef} f(\mathbf{u}) \, \sbelInfVolumeRef 
    =
    \int_{\sbelVolumeParent} f(\mathbf{u}(\boldsymbol{\xi})) \, J_p(\boldsymbol{\xi}) \, \sbelInfVolumeParent
    \approx \sum_{k=1}^{n_q} w_k \, f(\mathbf{u}(\boldsymbol{\xi}^q_k)) \, J_p(\boldsymbol{\xi}^q_k) \; ,
\end{equation}
where $\boldsymbol{\xi}^q_k$ are the quadrature points in the parent element, $w_k$ are the associated quadrature weights, and $n_q$ is the number of quadrature points. The quadrature points and weights are tabulated once on the parent element and reused for all elements.

The discussion above assumes an isoparametric element. If, instead, the reference element is obtained from the parent element by an affine transformation,
\[
\mathbf{u}(\boldsymbol{\xi}) = \mathbf{A}\boldsymbol{\xi} + \mathbf{b} \; ,
\]
with constant matrix $\mathbf A$ and constant vector $\mathbf b$, then
\[
\frac{\partial \mathbf{u}}{\partial \boldsymbol{\xi}} = \mathbf{A}
\qquad \text{and} \qquad
J_p(\boldsymbol{\xi}) = \det(\mathbf{A}) \; .
\]
The corresponding quadrature formula reduces to
\begin{equation}
\int_{\sbelVolumeRef} f(\mathbf{u}) \, \sbelInfVolumeRef 
=
\int_{\sbelVolumeParent} f(\mathbf{u}(\boldsymbol{\xi})) \, \det(\mathbf{A}) \, \sbelInfVolumeParent
\approx \det(\mathbf{A}) \sum_{k=1}^{n_q} w_k \, f(\mathbf{u}(\boldsymbol{\xi}^q_k)) \; .
\end{equation}
\end{subequations}

\section{Augmented Lagrangian Formulation}
\label{sec:alm_aspects}
The backward-Euler, constraint-enforced TL-FEA step is solved here as a velocity-level augmented-Lagrangian optimization problem. The idea is standard: after time discretization, the unknown configuration update is parameterized by the step velocity, the bilateral constraints are enforced through an augmented-Lagrangian term, and the discrete equations of motion are recovered as the stationarity conditions of the resulting objective; see, e.g., \cite{ortiz1999variational,hestenes1969multiplier,bertsekas2014constrained,opti:jorge}.

Under backward Euler, the step map is
\begin{equation}
\label{eq:step_map_velocity_level}
\mathbf{q}(\mathbf{v}) \coloneqq \mathbf{q}_n + h\,\mathbf{v},
\end{equation}
where $\mathbf{v}\equiv \mathbf{v}_{n+1}$ is the unknown velocity at time $t_{n+1}$. The position-level bilateral constraints are then evaluated through this map:
\begin{equation}
\label{eq:velocity_level_constraints}
\mathbf{c}\bigl(\mathbf{q}(\mathbf{v}),t_{n+1}\bigr)=\mathbf{0},
\qquad
\mathbf{C}_q\bigl(\mathbf{q}(\mathbf{v}),t_{n+1}\bigr)\coloneqq
\frac{\partial \mathbf{c}}{\partial \mathbf{q}}\bigl(\mathbf{q}(\mathbf{v}),t_{n+1}\bigr).
\end{equation}
Thus, although the constraints are posed at the position level, the unknown of the nonlinear solve is the velocity $\mathbf{v}$.

Using the force contributions derived in Sections~\ref{sec:dynamics_aspects} and~\ref{sec:frictional_contact_aspects}, the residual of the backward-Euler step is
\begin{equation}
\label{eq:residual}
\begin{aligned}
\mathbf{g}(\mathbf{v},\boldsymbol{\lambda})
&=
\frac{1}{h}\mathbf{M}(\mathbf{v}-\mathbf{v}_n)
+\mathbf{f}_{\mathrm{int}}\bigl(\mathbf{q}(\mathbf{v}),\mathbf{v}\bigr)
-\mathbf{f}_{\mathrm{ext}}
-\mathbf{f}_{\mathrm{ff}} \\
&\quad
+h\,\mathbf{C}_q\bigl(\mathbf{q}(\mathbf{v}),t_{n+1}\bigr)^{T}
\Bigl(
\boldsymbol{\lambda}
+\rho\,\mathbf{c}\bigl(\mathbf{q}(\mathbf{v}),t_{n+1}\bigr)
\Bigr) \; ,
\end{aligned}
\end{equation}
where $\mathbf{M}$ is the preassembled mass matrix, $\mathbf{f}_{\mathrm{int}}$ is the assembled internal-force vector, $\mathbf{f}_{\mathrm{ext}}$ collects externally applied nodal forces, $\mathbf{f}_{\mathrm{ff}}$ collects mass-distributed force-field contributions, $\boldsymbol{\lambda}\in\mathbb{R}^{m}$ is the vector of Lagrange multipliers, and $\rho>0$ is the augmented-Lagrangian penalty parameter. The constraint vector $\mathbf{c}\in\mathbb{R}^{m}$ lives in constraint space and is mapped back to generalized coordinates through $\mathbf{C}_q^{T}$.

The optimization interpretation is obtained by introducing an augmented objective in the velocity unknown. Assume that the internal-force operator admits an incremental potential $\Pi_{\mathrm{int}}(\mathbf{q},\mathbf{v})$ such that
\begin{equation}
\label{eq:internal_incremental_potential_assumption}
\nabla_{\mathbf{v}}
\left(
\frac{1}{h}\Pi_{\mathrm{int}}(\mathbf{q}(\mathbf{v}),\mathbf{v})
\right)
=
\mathbf{f}_{\mathrm{int}}\bigl(\mathbf{q}(\mathbf{v}),\mathbf{v}\bigr).
\end{equation}
For purely hyperelastic models, this follows from the stored-energy density. For Kelvin--Voigt-type models, the same structure is obtained by augmenting the elastic incremental potential with the corresponding discrete dissipation potential. Under this assumption, define
\begin{equation}
\label{eq:alm_cost_from_residual}
\begin{aligned}
\Phi_{\rho}(\mathbf{v},\boldsymbol{\lambda})
&=
\frac{1}{2h}\,
(\mathbf{v}-\mathbf{v}_n)^{T}\mathbf{M}(\mathbf{v}-\mathbf{v}_n)
+
\frac{1}{h}\,
\Pi_{\mathrm{int}}\bigl(\mathbf{q}(\mathbf{v}),\mathbf{v}\bigr) \\
&\quad
-\mathbf{f}_{\mathrm{ext}}^{T}\mathbf{v}
-\mathbf{f}_{\mathrm{ff}}^{T}\mathbf{v}
+\boldsymbol{\lambda}^{T}\mathbf{c}\bigl(\mathbf{q}(\mathbf{v}),t_{n+1}\bigr) \\
&\quad
+\frac{\rho}{2}\,
\left\|
\mathbf{c}\bigl(\mathbf{q}(\mathbf{v}),t_{n+1}\bigr)
\right\|_2^{2}.
\end{aligned}
\end{equation}
Using the chain rule and
\[
\nabla_{\mathbf{v}}\mathbf{c}\bigl(\mathbf{q}(\mathbf{v}),t_{n+1}\bigr)
=
h\,\mathbf{C}_q\bigl(\mathbf{q}(\mathbf{v}),t_{n+1}\bigr),
\]
one obtains
\begin{equation}
\label{eq:gradient_equals_residual}
\nabla_{\mathbf{v}}\Phi_{\rho}(\mathbf{v},\boldsymbol{\lambda})
=
\mathbf{g}(\mathbf{v},\boldsymbol{\lambda}).
\end{equation}
Therefore, for fixed multipliers $\boldsymbol{\lambda}$, the time-discrete constrained dynamics step is recast as the search for a stationary point of the augmented objective Eq.~\eqref{eq:alm_cost_from_residual}, and the discrete equations of motion are recovered from the first-order optimality condition
\[
\mathbf{g}(\mathbf{v},\boldsymbol{\lambda})=\mathbf{0}.
\]

The augmented-Lagrangian iteration then alternates between an inner solve for $\mathbf{v}$ at fixed $\boldsymbol{\lambda}$ and an outer multiplier update. In its simplest form, the multiplier update reads
\begin{equation}
\label{eq:alm_multiplier_update}
\boldsymbol{\lambda}^{(k+1)}
=
\boldsymbol{\lambda}^{(k)}
+
\rho\,
\mathbf{c}\bigl(\mathbf{q}(\mathbf{v}^{(k+1)}),t_{n+1}\bigr),
\end{equation}
and the outer loop is terminated once the constraint violation
\[
\left\|
\mathbf{c}\bigl(\mathbf{q}(\mathbf{v}^{(k+1)}),t_{n+1}\bigr)
\right\|_2
\]
falls below the prescribed tolerance. In the implementation, the transpose action of $\mathbf{C}_q$ is evaluated through the sparse element-level constraint blocks derived in Section~\ref{subsec:kinematic_constraints}, rather than through the explicit formation of a dense global matrix.

The discussion above is specialized to the purely deformable TL-FEA setting of this paper, in which the step unknown is the nodal velocity vector $\mathbf{v}$. The same velocity-level augmented-Lagrangian structure extends directly to mixed rigid/deformable systems by enlarging the optimization variable to include the translational and angular velocities of the rigid bodies and by evaluating the corresponding position-level constraints through the appropriate time-discrete step map.

\section{Conclusions}
\label{sec:conclusions}
This paper presented Part~I of a two-part contribution on Total Lagrangian finite element multibody dynamics. The emphasis here has been on formulation rather than performance: the goal was to define a consistent kinematic, constitutive, loading, contact, constraint, and time-discrete solution framework for systems of interacting deformable bodies undergoing finite deformation.

The main outcomes of the present work are as follows:
\begin{itemize}
  \item We developed a TL deformable-body dynamics formulation in which the kinematics and constitutive response are written consistently in terms of the deformation gradient $\mathbf F$ and its linearization, while adopting the standard structure of finite-strain continuum mechanics~\cite{Bonet2008nonlinear,Shabana1998}.
  \item By expressing the position field as $\mathbf r(\mathbf u;t)=\mathbf N(t)\mathbf s(\mathbf u)$, the deformation gradient admits the factorization $\mathbf F(\mathbf u;t)=\mathbf N(t)\mathbf H(\mathbf u)$. This separates the time-dependent nodal unknowns from precomputable, element-specific geometric quantities and provides a compact notation that carries through the derivation of internal forces, constraint Jacobians, and consistent linearizations~\cite{shabanaYakoub2001Part1,Shabana2023Overview}.
  \item We formulated deformable-body kinematic joints by composing scalar primitives of DP1, DP2, CD, and DIST type directly on isoparametric finite elements. Within this setting, the paper derived the associated element-level Jacobian blocks, discussed their accumulation into composite-joint operators, and identified the main linearization issues associated with dot-product constraints, including row scaling and the contribution of constraint curvature to the Newton system.
  \item We expressed the constitutive-discretization interface through the first Piola--Kirchhoff stress $\mathbf P(\mathbf F)$ and its material derivative $\partial \mathbf P/\partial \mathbf F$, which makes the internal-force and tangent constructions largely independent of element topology. In this form, the addition of a new hyperelastic or viscoelastic material model amounts to specifying $\mathbf P(\mathbf F)$ and its derivative, while the surrounding integration and assembly machinery remains unchanged~\cite{Gerstmayr2013,Otsuka2022RecentAdvances}.
  \item We derived, within one virtual-work framework, the contributions of inertia, mass-distributed force fields, internal forces, concentrated loads, prescribed surface tractions, frictional contact forces, and bilateral constraint reactions. The corresponding parent-to-reference quadrature construction was written in a form compatible with isoparametric finite elements and with the deformation-gradient-based constitutive interface.
  \item We cast the backward-Euler constrained TL-FEA step as a velocity-level augmented-Lagrangian optimization problem, in which the stationarity conditions recover the discrete equations of motion while bilateral constraints are enforced through multiplier and penalty terms. This provides a formulation-level link between the mechanics model and the nonlinear solution procedure used in practice.
\end{itemize}

The contribution of Part~I is therefore methodological: it establishes the formulation layer needed to model collections of deformable bodies, couple them through engineering joints, and advance them in time under finite deformation, external loading, contact, and constraints. Claims regarding computational efficiency, robustness, and large-scale numerical performance are intentionally deferred. Part~II implements the present formulation on the GPU and reports the resulting performance, numerical behavior, and benchmark results.

\backmatter

\section*{Statements and Declarations}

\begin{itemize}
\item \textbf{Funding}\enspace This work was supported in part by the
National Science Foundation under grant OAC2209791.
\item \textbf{Competing Interests}\enspace The authors declare no competing
interests.
\item \textbf{Data availability}\enspace Not applicable.
\end{itemize}

\begin{appendices}

\section{Select Element Types}
\label{sec:select_element_types}

\subsection{10-Node Tetrahedron Element}
\label{subsubsec:tet10-canonical-tetrahedron-element}
The Tet10 Canonical element has the following characteristics:
\begin{itemize}
    \item Element type: 10-Node Tetrahedron (Tet10-P)
    \item Number of nodes: 10
    \item Number of nodal unknowns per node: 1 (position vector $\mathbf{r}$)
    \item Total number of nodal unknowns: 10
    \item Basis: Quadratic Lagrange polynomials in barycentric coordinates
\end{itemize}

\paragraph{Node Locations}
The nodal locations in the reference coordinate system are:
\begin{verbatim}
Node # | (L1, L2, L3, L4) | (xi, eta, zeta)
---------------------------------------------
1      | (   1,    0,    0,    0) | (   0,    0,    0)
2      | (   0,    1,    0,    0) | (   1,    0,    0)
3      | (   0,    0,    1,    0) | (   0,    1,    0)
4      | (   0,    0,    0,    1) | (   0,    0,    1)
5      | ( 1/2,  1/2,    0,    0) | ( 1/2,    0,    0)
6      | (   0,  1/2,  1/2,    0) | ( 1/2,  1/2,    0)
7      | ( 1/2,    0,  1/2,    0) | (   0,  1/2,    0)
8      | ( 1/2,    0,    0,  1/2) | (   0,    0,  1/2)
9      | (   0,  1/2,    0,  1/2) | ( 1/2,    0,  1/2)
10     | (   0,    0,  1/2,  1/2) | (   0,  1/2,  1/2)
\end{verbatim}

\paragraph{Barycentric Mapping}
We use barycentric coordinates $L_1, L_2, L_3, L_4$ on the tetrahedron with canonical coordinates $(\xi, \eta, \zeta)$ defined by
\[
\begin{aligned}
L_1 &= 1 - \xi - \eta - \zeta, \\
L_2 &= \xi, \\
L_3 &= \eta, \\
L_4 &= \zeta.
\end{aligned}
\]
The inverse mapping is immediate: $\xi = L_2$, $\eta = L_3$, $\zeta = L_4$, with the constraints $L_i \ge 0$ and $L_1 + L_2 + L_3 + L_4 = 1$.

\paragraph{Shape Functions}
The resulting vector of 10 Lagrange shape functions, $\mathbf{s}(\xi, \eta, \zeta)$, is:
\[ \mathbf{s}(\xi, \eta, \zeta) = \left[\begin{matrix}2 \eta^{2} + 4 \eta \xi + 4 \eta \zeta - 3 \eta + 2 \xi^{2} + 4 \xi \zeta - 3 \xi + 2 \zeta^{2} - 3 \zeta + 1\\2 \xi^{2} - \xi\\2 \eta^{2} - \eta\\2 \zeta^{2} - \zeta\\- 4 \eta \xi - 4 \xi^{2} - 4 \xi \zeta + 4 \xi\\4 \eta \xi\\- 4 \eta^{2} - 4 \eta \xi - 4 \eta \zeta + 4 \eta\\- 4 \eta \zeta - 4 \xi \zeta - 4 \zeta^{2} + 4 \zeta\\4 \xi \zeta\\4 \eta \zeta\end{matrix}\right] \]

\paragraph{Shape Function Derivatives}
The Jacobian of the shape functions with respect to the canonical element coordinates is:
\[ \mathbf{H} = \left[\begin{matrix}4 \eta + 4 \xi + 4 \zeta - 3 & 4 \eta + 4 \xi + 4 \zeta - 3 & 4 \eta + 4 \xi + 4 \zeta - 3\\4 \xi - 1 & 0 & 0\\0 & 4 \eta - 1 & 0\\0 & 0 & 4 \zeta - 1\\- 4 \eta - 8 \xi - 4 \zeta + 4 & - 4 \xi & - 4 \xi\\4 \eta & 4 \xi & 0\\- 4 \eta & - 8 \eta - 4 \xi - 4 \zeta + 4 & - 4 \eta\\- 4 \zeta & - 4 \zeta & - 4 \eta - 4 \xi - 8 \zeta + 4\\4 \zeta & 0 & 4 \xi\\0 & 4 \zeta & 4 \eta\end{matrix}\right] \]


\subsection{The Q27 hexahedron element}
\label{subsubsec:hex27-canonical-hexahedron-element}
The Q27 element has the following characteristics:
\begin{itemize}
    \item Element type: 27-Node Canonical Hexahedron Element (Q27-P)
    \item Number of nodes: 27
    \item Number of nodal unknowns per node: 1 (position vector $\mathbf{r}$)
    \item Total number of nodal unknowns: 27
    \item Basis: Tensor product of 1D quadratic Lagrange polynomials
\end{itemize}

\paragraph{Node Locations}
The nodal locations in the reference coordinate system are:
\begin{verbatim}
Node # | (i,j,k) | (xi, eta, zeta)
----------------------------------
1      | (0,0,0)     | (-1, -1, -1)
2      | (1,0,0)     | ( 0, -1, -1)
3      | (2,0,0)     | ( 1, -1, -1)
4      | (0,1,0)     | (-1,  0, -1)
5      | (1,1,0)     | ( 0,  0, -1)
6      | (2,1,0)     | ( 1,  0, -1)
7      | (0,2,0)     | (-1,  1, -1)
8      | (1,2,0)     | ( 0,  1, -1)
9      | (2,2,0)     | ( 1,  1, -1)
10     | (0,0,1)     | (-1, -1,  0)
11     | (1,0,1)     | ( 0, -1,  0)
12     | (2,0,1)     | ( 1, -1,  0)
13     | (0,1,1)     | (-1,  0,  0)
14     | (1,1,1)     | ( 0,  0,  0)
15     | (2,1,1)     | ( 1,  0,  0)
16     | (0,2,1)     | (-1,  1,  0)
17     | (1,2,1)     | ( 0,  1,  0)
18     | (2,2,1)     | ( 1,  1,  0)
19     | (0,0,2)     | (-1, -1,  1)
20     | (1,0,2)     | ( 0, -1,  1)
21     | (2,0,2)     | ( 1, -1,  1)
22     | (0,1,2)     | (-1,  0,  1)
23     | (1,1,2)     | ( 0,  0,  1)
24     | (2,1,2)     | ( 1,  0,  1)
25     | (0,2,2)     | (-1,  1,  1)
26     | (1,2,2)     | ( 0,  1,  1)
27     | (2,2,2)     | ( 1,  1,  1)
\end{verbatim}

\paragraph{1D Lagrange Polynomials}
The 1D quadratic Lagrange polynomials used as the basis are:
\[ \mathbf{L}(s) = \left[\begin{matrix}\frac{s \left(s - 1\right)}{2}\\1 - s^{2}\\\frac{s \left(s + 1\right)}{2}\end{matrix}\right] \]

\paragraph{Shape Functions}
The resulting vector of 27 shape functions, $\mathbf{s}(\xi, \eta, \zeta)$, are tensor products of the 1D Lagrange polynomials. The expressions are:
\[ \mathbf{s}(\xi, \eta, \zeta) = \left[\begin{matrix}\frac{\eta^{2} \xi^{2} \zeta^{2}}{8} - \frac{\eta^{2} \xi^{2} \zeta}{8} - \frac{\eta^{2} \xi \zeta^{2}}{8} + \frac{\eta^{2} \xi \zeta}{8} - \frac{\eta \xi^{2} \zeta^{2}}{8} + \frac{\eta \xi^{2} \zeta}{8} + \frac{\eta \xi \zeta^{2}}{8} - \frac{\eta \xi \zeta}{8}\\- \frac{\eta^{2} \xi^{2} \zeta^{2}}{4} + \frac{\eta^{2} \xi^{2} \zeta}{4} + \frac{\eta^{2} \zeta^{2}}{4} - \frac{\eta^{2} \zeta}{4} + \frac{\eta \xi^{2} \zeta^{2}}{4} - \frac{\eta \xi^{2} \zeta}{4} - \frac{\eta \zeta^{2}}{4} + \frac{\eta \zeta}{4}\\\frac{\eta^{2} \xi^{2} \zeta^{2}}{8} - \frac{\eta^{2} \xi^{2} \zeta}{8} + \frac{\eta^{2} \xi \zeta^{2}}{8} - \frac{\eta^{2} \xi \zeta}{8} - \frac{\eta \xi^{2} \zeta^{2}}{8} + \frac{\eta \xi^{2} \zeta}{8} - \frac{\eta \xi \zeta^{2}}{8} + \frac{\eta \xi \zeta}{8}\\- \frac{\eta^{2} \xi^{2} \zeta^{2}}{4} + \frac{\eta^{2} \xi^{2} \zeta}{4} + \frac{\eta^{2} \xi \zeta^{2}}{4} - \frac{\eta^{2} \xi \zeta}{4} + \frac{\xi^{2} \zeta^{2}}{4} - \frac{\xi^{2} \zeta}{4} - \frac{\xi \zeta^{2}}{4} + \frac{\xi \zeta}{4}\\\frac{\eta^{2} \xi^{2} \zeta^{2}}{2} - \frac{\eta^{2} \xi^{2} \zeta}{2} - \frac{\eta^{2} \zeta^{2}}{2} + \frac{\eta^{2} \zeta}{2} - \frac{\xi^{2} \zeta^{2}}{2} + \frac{\xi^{2} \zeta}{2} + \frac{\zeta^{2}}{2} - \frac{\zeta}{2}\\- \frac{\eta^{2} \xi^{2} \zeta^{2}}{4} + \frac{\eta^{2} \xi^{2} \zeta}{4} - \frac{\eta^{2} \xi \zeta^{2}}{4} + \frac{\eta^{2} \xi \zeta}{4} + \frac{\xi^{2} \zeta^{2}}{4} - \frac{\xi^{2} \zeta}{4} + \frac{\xi \zeta^{2}}{4} - \frac{\xi \zeta}{4}\\\frac{\eta^{2} \xi^{2} \zeta^{2}}{8} - \frac{\eta^{2} \xi^{2} \zeta}{8} - \frac{\eta^{2} \xi \zeta^{2}}{8} + \frac{\eta^{2} \xi \zeta}{8} + \frac{\eta \xi^{2} \zeta^{2}}{8} - \frac{\eta \xi^{2} \zeta}{8} - \frac{\eta \xi \zeta^{2}}{8} + \frac{\eta \xi \zeta}{8}\\- \frac{\eta^{2} \xi^{2} \zeta^{2}}{4} + \frac{\eta^{2} \xi^{2} \zeta}{4} + \frac{\eta^{2} \zeta^{2}}{4} - \frac{\eta^{2} \zeta}{4} - \frac{\eta \xi^{2} \zeta^{2}}{4} + \frac{\eta \xi^{2} \zeta}{4} + \frac{\eta \zeta^{2}}{4} - \frac{\eta \zeta}{4}\\\frac{\eta^{2} \xi^{2} \zeta^{2}}{8} - \frac{\eta^{2} \xi^{2} \zeta}{8} + \frac{\eta^{2} \xi \zeta^{2}}{8} - \frac{\eta^{2} \xi \zeta}{8} + \frac{\eta \xi^{2} \zeta^{2}}{8} - \frac{\eta \xi^{2} \zeta}{8} + \frac{\eta \xi \zeta^{2}}{8} - \frac{\eta \xi \zeta}{8}\\- \frac{\eta^{2} \xi^{2} \zeta^{2}}{4} + \frac{\eta^{2} \xi^{2}}{4} + \frac{\eta^{2} \xi \zeta^{2}}{4} - \frac{\eta^{2} \xi}{4} + \frac{\eta \xi^{2} \zeta^{2}}{4} - \frac{\eta \xi^{2}}{4} - \frac{\eta \xi \zeta^{2}}{4} + \frac{\eta \xi}{4}\\\frac{\eta^{2} \xi^{2} \zeta^{2}}{2} - \frac{\eta^{2} \xi^{2}}{2} - \frac{\eta^{2} \zeta^{2}}{2} + \frac{\eta^{2}}{2} - \frac{\eta \xi^{2} \zeta^{2}}{2} + \frac{\eta \xi^{2}}{2} + \frac{\eta \zeta^{2}}{2} - \frac{\eta}{2}\\- \frac{\eta^{2} \xi^{2} \zeta^{2}}{4} + \frac{\eta^{2} \xi^{2}}{4} - \frac{\eta^{2} \xi \zeta^{2}}{4} + \frac{\eta^{2} \xi}{4} + \frac{\eta \xi^{2} \zeta^{2}}{4} - \frac{\eta \xi^{2}}{4} + \frac{\eta \xi \zeta^{2}}{4} - \frac{\eta \xi}{4}\\\frac{\eta^{2} \xi^{2} \zeta^{2}}{2} - \frac{\eta^{2} \xi^{2}}{2} - \frac{\eta^{2} \xi \zeta^{2}}{2} + \frac{\eta^{2} \xi}{2} - \frac{\xi^{2} \zeta^{2}}{2} + \frac{\xi^{2}}{2} + \frac{\xi \zeta^{2}}{2} - \frac{\xi}{2}\\- \eta^{2} \xi^{2} \zeta^{2} + \eta^{2} \xi^{2} + \eta^{2} \zeta^{2} - \eta^{2} + \xi^{2} \zeta^{2} - \xi^{2} - \zeta^{2} + 1\\\frac{\eta^{2} \xi^{2} \zeta^{2}}{2} - \frac{\eta^{2} \xi^{2}}{2} + \frac{\eta^{2} \xi \zeta^{2}}{2} - \frac{\eta^{2} \xi}{2} - \frac{\xi^{2} \zeta^{2}}{2} + \frac{\xi^{2}}{2} - \frac{\xi \zeta^{2}}{2} + \frac{\xi}{2}\\- \frac{\eta^{2} \xi^{2} \zeta^{2}}{4} + \frac{\eta^{2} \xi^{2}}{4} + \frac{\eta^{2} \xi \zeta^{2}}{4} - \frac{\eta^{2} \xi}{4} - \frac{\eta \xi^{2} \zeta^{2}}{4} + \frac{\eta \xi^{2}}{4} + \frac{\eta \xi \zeta^{2}}{4} - \frac{\eta \xi}{4}\\\frac{\eta^{2} \xi^{2} \zeta^{2}}{2} - \frac{\eta^{2} \xi^{2}}{2} - \frac{\eta^{2} \zeta^{2}}{2} + \frac{\eta^{2}}{2} + \frac{\eta \xi^{2} \zeta^{2}}{2} - \frac{\eta \xi^{2}}{2} - \frac{\eta \zeta^{2}}{2} + \frac{\eta}{2}\\- \frac{\eta^{2} \xi^{2} \zeta^{2}}{4} + \frac{\eta^{2} \xi^{2}}{4} - \frac{\eta^{2} \xi \zeta^{2}}{4} + \frac{\eta^{2} \xi}{4} - \frac{\eta \xi^{2} \zeta^{2}}{4} + \frac{\eta \xi^{2}}{4} - \frac{\eta \xi \zeta^{2}}{4} + \frac{\eta \xi}{4}\\\frac{\eta^{2} \xi^{2} \zeta^{2}}{8} + \frac{\eta^{2} \xi^{2} \zeta}{8} - \frac{\eta^{2} \xi \zeta^{2}}{8} - \frac{\eta^{2} \xi \zeta}{8} - \frac{\eta \xi^{2} \zeta^{2}}{8} - \frac{\eta \xi^{2} \zeta}{8} + \frac{\eta \xi \zeta^{2}}{8} + \frac{\eta \xi \zeta}{8}\\- \frac{\eta^{2} \xi^{2} \zeta^{2}}{4} - \frac{\eta^{2} \xi^{2} \zeta}{4} + \frac{\eta^{2} \zeta^{2}}{4} + \frac{\eta^{2} \zeta}{4} + \frac{\eta \xi^{2} \zeta^{2}}{4} + \frac{\eta \xi^{2} \zeta}{4} - \frac{\eta \zeta^{2}}{4} - \frac{\eta \zeta}{4}\\\frac{\eta^{2} \xi^{2} \zeta^{2}}{8} + \frac{\eta^{2} \xi^{2} \zeta}{8} + \frac{\eta^{2} \xi \zeta^{2}}{8} + \frac{\eta^{2} \xi \zeta}{8} - \frac{\eta \xi^{2} \zeta^{2}}{8} - \frac{\eta \xi^{2} \zeta}{8} - \frac{\eta \xi \zeta^{2}}{8} - \frac{\eta \xi \zeta}{8}\\- \frac{\eta^{2} \xi^{2} \zeta^{2}}{4} - \frac{\eta^{2} \xi^{2} \zeta}{4} + \frac{\eta^{2} \xi \zeta^{2}}{4} + \frac{\eta^{2} \xi \zeta}{4} + \frac{\xi^{2} \zeta^{2}}{4} + \frac{\xi^{2} \zeta}{4} - \frac{\xi \zeta^{2}}{4} - \frac{\xi \zeta}{4}\\\frac{\eta^{2} \xi^{2} \zeta^{2}}{2} + \frac{\eta^{2} \xi^{2} \zeta}{2} - \frac{\eta^{2} \zeta^{2}}{2} - \frac{\eta^{2} \zeta}{2} - \frac{\xi^{2} \zeta^{2}}{2} - \frac{\xi^{2} \zeta}{2} + \frac{\zeta^{2}}{2} + \frac{\zeta}{2}\\- \frac{\eta^{2} \xi^{2} \zeta^{2}}{4} - \frac{\eta^{2} \xi^{2} \zeta}{4} - \frac{\eta^{2} \xi \zeta^{2}}{4} - \frac{\eta^{2} \xi \zeta}{4} + \frac{\xi^{2} \zeta^{2}}{4} + \frac{\xi^{2} \zeta}{4} + \frac{\xi \zeta^{2}}{4} + \frac{\xi \zeta}{4}\\\frac{\eta^{2} \xi^{2} \zeta^{2}}{8} + \frac{\eta^{2} \xi^{2} \zeta}{8} - \frac{\eta^{2} \xi \zeta^{2}}{8} - \frac{\eta^{2} \xi \zeta}{8} + \frac{\eta \xi^{2} \zeta^{2}}{8} + \frac{\eta \xi^{2} \zeta}{8} - \frac{\eta \xi \zeta^{2}}{8} - \frac{\eta \xi \zeta}{8}\\- \frac{\eta^{2} \xi^{2} \zeta^{2}}{4} - \frac{\eta^{2} \xi^{2} \zeta}{4} + \frac{\eta^{2} \zeta^{2}}{4} + \frac{\eta^{2} \zeta}{4} - \frac{\eta \xi^{2} \zeta^{2}}{4} - \frac{\eta \xi^{2} \zeta}{4} + \frac{\eta \zeta^{2}}{4} + \frac{\eta \zeta}{4}\\\frac{\eta^{2} \xi^{2} \zeta^{2}}{8} + \frac{\eta^{2} \xi^{2} \zeta}{8} + \frac{\eta^{2} \xi \zeta^{2}}{8} + \frac{\eta^{2} \xi \zeta}{8} + \frac{\eta \xi^{2} \zeta^{2}}{8} + \frac{\eta \xi^{2} \zeta}{8} + \frac{\eta \xi \zeta^{2}}{8} + \frac{\eta \xi \zeta}{8}\end{matrix}\right] \]

\paragraph{Shape Function Derivatives}
The Jacobian of the shape functions with respect to the canonical coordinates is:
\[ \mathbf{H} = \resizebox{\linewidth}{!}{$\left[\begin{matrix}\frac{\eta \zeta \left(2 \eta \xi \zeta - 2 \eta \xi - \eta \zeta + \eta - 2 \xi \zeta + 2 \xi + \zeta - 1\right)}{8} & \frac{\xi \zeta \left(2 \eta \xi \zeta - 2 \eta \xi - 2 \eta \zeta + 2 \eta - \xi \zeta + \xi + \zeta - 1\right)}{8} & \frac{\eta \xi \left(2 \eta \xi \zeta - \eta \xi - 2 \eta \zeta + \eta - 2 \xi \zeta + \xi + 2 \zeta - 1\right)}{8}\\\frac{\eta \xi \zeta \left(- \eta \zeta + \eta + \zeta - 1\right)}{2} & \frac{\zeta \left(- 2 \eta \xi^{2} \zeta + 2 \eta \xi^{2} + 2 \eta \zeta - 2 \eta + \xi^{2} \zeta - \xi^{2} - \zeta + 1\right)}{4} & \frac{\eta \left(- 2 \eta \xi^{2} \zeta + \eta \xi^{2} + 2 \eta \zeta - \eta + 2 \xi^{2} \zeta - \xi^{2} - 2 \zeta + 1\right)}{4}\\\frac{\eta \zeta \left(2 \eta \xi \zeta - 2 \eta \xi + \eta \zeta - \eta - 2 \xi \zeta + 2 \xi - \zeta + 1\right)}{8} & \frac{\xi \zeta \left(2 \eta \xi \zeta - 2 \eta \xi + 2 \eta \zeta - 2 \eta - \xi \zeta + \xi - \zeta + 1\right)}{8} & \frac{\eta \xi \left(2 \eta \xi \zeta - \eta \xi + 2 \eta \zeta - \eta - 2 \xi \zeta + \xi - 2 \zeta + 1\right)}{8}\\\frac{\zeta \left(- 2 \eta^{2} \xi \zeta + 2 \eta^{2} \xi + \eta^{2} \zeta - \eta^{2} + 2 \xi \zeta - 2 \xi - \zeta + 1\right)}{4} & \frac{\eta \xi \zeta \left(- \xi \zeta + \xi + \zeta - 1\right)}{2} & \frac{\xi \left(- 2 \eta^{2} \xi \zeta + \eta^{2} \xi + 2 \eta^{2} \zeta - \eta^{2} + 2 \xi \zeta - \xi - 2 \zeta + 1\right)}{4}\\\xi \zeta \left(\eta^{2} \zeta - \eta^{2} - \zeta + 1\right) & \eta \zeta \left(\xi^{2} \zeta - \xi^{2} - \zeta + 1\right) & \eta^{2} \xi^{2} \zeta - \frac{\eta^{2} \xi^{2}}{2} - \eta^{2} \zeta + \frac{\eta^{2}}{2} - \xi^{2} \zeta + \frac{\xi^{2}}{2} + \zeta - \frac{1}{2}\\\frac{\zeta \left(- 2 \eta^{2} \xi \zeta + 2 \eta^{2} \xi - \eta^{2} \zeta + \eta^{2} + 2 \xi \zeta - 2 \xi + \zeta - 1\right)}{4} & \frac{\eta \xi \zeta \left(- \xi \zeta + \xi - \zeta + 1\right)}{2} & \frac{\xi \left(- 2 \eta^{2} \xi \zeta + \eta^{2} \xi - 2 \eta^{2} \zeta + \eta^{2} + 2 \xi \zeta - \xi + 2 \zeta - 1\right)}{4}\\\frac{\eta \zeta \left(2 \eta \xi \zeta - 2 \eta \xi - \eta \zeta + \eta + 2 \xi \zeta - 2 \xi - \zeta + 1\right)}{8} & \frac{\xi \zeta \left(2 \eta \xi \zeta - 2 \eta \xi - 2 \eta \zeta + 2 \eta + \xi \zeta - \xi - \zeta + 1\right)}{8} & \frac{\eta \xi \left(2 \eta \xi \zeta - \eta \xi - 2 \eta \zeta + \eta + 2 \xi \zeta - \xi - 2 \zeta + 1\right)}{8}\\\frac{\eta \xi \zeta \left(- \eta \zeta + \eta - \zeta + 1\right)}{2} & \frac{\zeta \left(- 2 \eta \xi^{2} \zeta + 2 \eta \xi^{2} + 2 \eta \zeta - 2 \eta - \xi^{2} \zeta + \xi^{2} + \zeta - 1\right)}{4} & \frac{\eta \left(- 2 \eta \xi^{2} \zeta + \eta \xi^{2} + 2 \eta \zeta - \eta - 2 \xi^{2} \zeta + \xi^{2} + 2 \zeta - 1\right)}{4}\\\frac{\eta \zeta \left(2 \eta \xi \zeta - 2 \eta \xi + \eta \zeta - \eta + 2 \xi \zeta - 2 \xi + \zeta - 1\right)}{8} & \frac{\xi \zeta \left(2 \eta \xi \zeta - 2 \eta \xi + 2 \eta \zeta - 2 \eta + \xi \zeta - \xi + \zeta - 1\right)}{8} & \frac{\eta \xi \left(2 \eta \xi \zeta - \eta \xi + 2 \eta \zeta - \eta + 2 \xi \zeta - \xi + 2 \zeta - 1\right)}{8}\\\frac{\eta \left(- 2 \eta \xi \zeta^{2} + 2 \eta \xi + \eta \zeta^{2} - \eta + 2 \xi \zeta^{2} - 2 \xi - \zeta^{2} + 1\right)}{4} & \frac{\xi \left(- 2 \eta \xi \zeta^{2} + 2 \eta \xi + 2 \eta \zeta^{2} - 2 \eta + \xi \zeta^{2} - \xi - \zeta^{2} + 1\right)}{4} & \frac{\eta \xi \zeta \left(- \eta \xi + \eta + \xi - 1\right)}{2}\\\eta \xi \left(\eta \zeta^{2} - \eta - \zeta^{2} + 1\right) & \eta \xi^{2} \zeta^{2} - \eta \xi^{2} - \eta \zeta^{2} + \eta - \frac{\xi^{2} \zeta^{2}}{2} + \frac{\xi^{2}}{2} + \frac{\zeta^{2}}{2} - \frac{1}{2} & \eta \zeta \left(\eta \xi^{2} - \eta - \xi^{2} + 1\right)\\\frac{\eta \left(- 2 \eta \xi \zeta^{2} + 2 \eta \xi - \eta \zeta^{2} + \eta + 2 \xi \zeta^{2} - 2 \xi + \zeta^{2} - 1\right)}{4} & \frac{\xi \left(- 2 \eta \xi \zeta^{2} + 2 \eta \xi - 2 \eta \zeta^{2} + 2 \eta + \xi \zeta^{2} - \xi + \zeta^{2} - 1\right)}{4} & \frac{\eta \xi \zeta \left(- \eta \xi - \eta + \xi + 1\right)}{2}\\\eta^{2} \xi \zeta^{2} - \eta^{2} \xi - \frac{\eta^{2} \zeta^{2}}{2} + \frac{\eta^{2}}{2} - \xi \zeta^{2} + \xi + \frac{\zeta^{2}}{2} - \frac{1}{2} & \eta \xi \left(\xi \zeta^{2} - \xi - \zeta^{2} + 1\right) & \xi \zeta \left(\eta^{2} \xi - \eta^{2} - \xi + 1\right)\\2 \xi \left(- \eta^{2} \zeta^{2} + \eta^{2} + \zeta^{2} - 1\right) & 2 \eta \left(- \xi^{2} \zeta^{2} + \xi^{2} + \zeta^{2} - 1\right) & 2 \zeta \left(- \eta^{2} \xi^{2} + \eta^{2} + \xi^{2} - 1\right)\\\eta^{2} \xi \zeta^{2} - \eta^{2} \xi + \frac{\eta^{2} \zeta^{2}}{2} - \frac{\eta^{2}}{2} - \xi \zeta^{2} + \xi - \frac{\zeta^{2}}{2} + \frac{1}{2} & \eta \xi \left(\xi \zeta^{2} - \xi + \zeta^{2} - 1\right) & \xi \zeta \left(\eta^{2} \xi + \eta^{2} - \xi - 1\right)\\\frac{\eta \left(- 2 \eta \xi \zeta^{2} + 2 \eta \xi + \eta \zeta^{2} - \eta - 2 \xi \zeta^{2} + 2 \xi + \zeta^{2} - 1\right)}{4} & \frac{\xi \left(- 2 \eta \xi \zeta^{2} + 2 \eta \xi + 2 \eta \zeta^{2} - 2 \eta - \xi \zeta^{2} + \xi + \zeta^{2} - 1\right)}{4} & \frac{\eta \xi \zeta \left(- \eta \xi + \eta - \xi + 1\right)}{2}\\\eta \xi \left(\eta \zeta^{2} - \eta + \zeta^{2} - 1\right) & \eta \xi^{2} \zeta^{2} - \eta \xi^{2} - \eta \zeta^{2} + \eta + \frac{\xi^{2} \zeta^{2}}{2} - \frac{\xi^{2}}{2} - \frac{\zeta^{2}}{2} + \frac{1}{2} & \eta \zeta \left(\eta \xi^{2} - \eta + \xi^{2} - 1\right)\\\frac{\eta \left(- 2 \eta \xi \zeta^{2} + 2 \eta \xi - \eta \zeta^{2} + \eta - 2 \xi \zeta^{2} + 2 \xi - \zeta^{2} + 1\right)}{4} & \frac{\xi \left(- 2 \eta \xi \zeta^{2} + 2 \eta \xi - 2 \eta \zeta^{2} + 2 \eta - \xi \zeta^{2} + \xi - \zeta^{2} + 1\right)}{4} & \frac{\eta \xi \zeta \left(- \eta \xi - \eta - \xi - 1\right)}{2}\\\frac{\eta \zeta \left(2 \eta \xi \zeta + 2 \eta \xi - \eta \zeta - \eta - 2 \xi \zeta - 2 \xi + \zeta + 1\right)}{8} & \frac{\xi \zeta \left(2 \eta \xi \zeta + 2 \eta \xi - 2 \eta \zeta - 2 \eta - \xi \zeta - \xi + \zeta + 1\right)}{8} & \frac{\eta \xi \left(2 \eta \xi \zeta + \eta \xi - 2 \eta \zeta - \eta - 2 \xi \zeta - \xi + 2 \zeta + 1\right)}{8}\\\frac{\eta \xi \zeta \left(- \eta \zeta - \eta + \zeta + 1\right)}{2} & \frac{\zeta \left(- 2 \eta \xi^{2} \zeta - 2 \eta \xi^{2} + 2 \eta \zeta + 2 \eta + \xi^{2} \zeta + \xi^{2} - \zeta - 1\right)}{4} & \frac{\eta \left(- 2 \eta \xi^{2} \zeta - \eta \xi^{2} + 2 \eta \zeta + \eta + 2 \xi^{2} \zeta + \xi^{2} - 2 \zeta - 1\right)}{4}\\\frac{\eta \zeta \left(2 \eta \xi \zeta + 2 \eta \xi + \eta \zeta + \eta - 2 \xi \zeta - 2 \xi - \zeta - 1\right)}{8} & \frac{\xi \zeta \left(2 \eta \xi \zeta + 2 \eta \xi + 2 \eta \zeta + 2 \eta - \xi \zeta - \xi - \zeta - 1\right)}{8} & \frac{\eta \xi \left(2 \eta \xi \zeta + \eta \xi + 2 \eta \zeta + \eta - 2 \xi \zeta - \xi - 2 \zeta - 1\right)}{8}\\\frac{\zeta \left(- 2 \eta^{2} \xi \zeta - 2 \eta^{2} \xi + \eta^{2} \zeta + \eta^{2} + 2 \xi \zeta + 2 \xi - \zeta - 1\right)}{4} & \frac{\eta \xi \zeta \left(- \xi \zeta - \xi + \zeta + 1\right)}{2} & \frac{\xi \left(- 2 \eta^{2} \xi \zeta - \eta^{2} \xi + 2 \eta^{2} \zeta + \eta^{2} + 2 \xi \zeta + \xi - 2 \zeta - 1\right)}{4}\\\xi \zeta \left(\eta^{2} \zeta + \eta^{2} - \zeta - 1\right) & \eta \zeta \left(\xi^{2} \zeta + \xi^{2} - \zeta - 1\right) & \eta^{2} \xi^{2} \zeta + \frac{\eta^{2} \xi^{2}}{2} - \eta^{2} \zeta - \frac{\eta^{2}}{2} - \xi^{2} \zeta - \frac{\xi^{2}}{2} + \zeta + \frac{1}{2}\\\frac{\zeta \left(- 2 \eta^{2} \xi \zeta - 2 \eta^{2} \xi - \eta^{2} \zeta - \eta^{2} + 2 \xi \zeta + 2 \xi + \zeta + 1\right)}{4} & \frac{\eta \xi \zeta \left(- \xi \zeta - \xi - \zeta - 1\right)}{2} & \frac{\xi \left(- 2 \eta^{2} \xi \zeta - \eta^{2} \xi - 2 \eta^{2} \zeta - \eta^{2} + 2 \xi \zeta + \xi + 2 \zeta + 1\right)}{4}\\\frac{\eta \zeta \left(2 \eta \xi \zeta + 2 \eta \xi - \eta \zeta - \eta + 2 \xi \zeta + 2 \xi - \zeta - 1\right)}{8} & \frac{\xi \zeta \left(2 \eta \xi \zeta + 2 \eta \xi - 2 \eta \zeta - 2 \eta + \xi \zeta + \xi - \zeta - 1\right)}{8} & \frac{\eta \xi \left(2 \eta \xi \zeta + \eta \xi - 2 \eta \zeta - \eta + 2 \xi \zeta + \xi - 2 \zeta - 1\right)}{8}\\\frac{\eta \xi \zeta \left(- \eta \zeta - \eta - \zeta - 1\right)}{2} & \frac{\zeta \left(- 2 \eta \xi^{2} \zeta - 2 \eta \xi^{2} + 2 \eta \zeta + 2 \eta - \xi^{2} \zeta - \xi^{2} + \zeta + 1\right)}{4} & \frac{\eta \left(- 2 \eta \xi^{2} \zeta - \eta \xi^{2} + 2 \eta \zeta + \eta - 2 \xi^{2} \zeta - \xi^{2} + 2 \zeta + 1\right)}{4}\\\frac{\eta \zeta \left(2 \eta \xi \zeta + 2 \eta \xi + \eta \zeta + \eta + 2 \xi \zeta + 2 \xi + \zeta + 1\right)}{8} & \frac{\xi \zeta \left(2 \eta \xi \zeta + 2 \eta \xi + 2 \eta \zeta + 2 \eta + \xi \zeta + \xi + \zeta + 1\right)}{8} & \frac{\eta \xi \left(2 \eta \xi \zeta + \eta \xi + 2 \eta \zeta + \eta + 2 \xi \zeta + \xi + 2 \zeta + 1\right)}{8}\end{matrix}\right]$} \]


\subsection{The 3243 ANCF Beam Element}
\label{subsec:3243-ancf-beam-element}
\begin{figure}[htb]
    \centering
    \safeincludegraphics[width=0.7\textwidth]{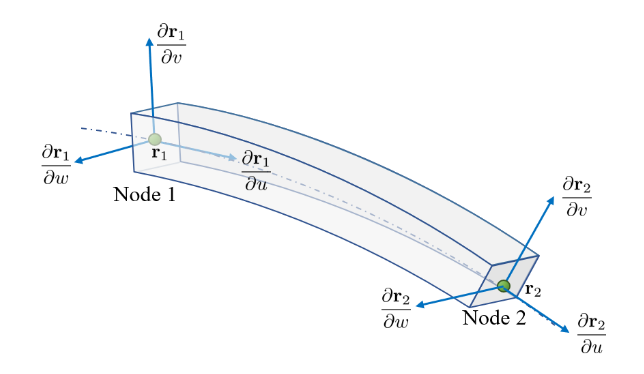}
    \caption{Visualization of a fully-parameterized 2-node ANCF beam element 3243}
    \label{fig:3243beam}
\end{figure}

The most basic element considered in the comparison is the original 2-node ANCF beam element 3243, which is fully parameterized~\cite{gerstmayrECCOMAS2005}. As illustrated in Fig.~\ref{fig:3243beam}, each node is described by 12 coordinates, including a position vector $\mathbf{r}_i$ and its gradients with respect to the local coordinates $u$, $v$, and $w$, denoted as $\frac{\partial \mathbf{r}_i}{\partial u}$, $\frac{\partial \mathbf{r}_i}{\partial v}$, and $\frac{\partial \mathbf{r}_i}{\partial w}$. This full set of coordinates per node defines the ``fully parameterized'' terminology. With two such nodes, the element possesses 24 degrees of freedom when unconstrained.

The vector of basis function for 3243 beam element is:
\[
\mathbf{b}^T(u,v,w) = [1, u, v, w, uv, uw, u^2, u^3]
\]

The constant matrix $\mathbf{B}_{12} \in \mathbb{R}^{8 \times 8}$ can be defined as:
\[
\begin{aligned}
\mathbf{B}_{12} 
\equiv [&\mathbf{b}(P_1),\;
\mathbf{b}_{,u}(P_1),\;
\mathbf{b}_{,v}(P_1),\;
\mathbf{b}_{,w}(P_1), \\
&\mathbf{b}(P_2),\;
\mathbf{b}_{,u}(P_2),\;
\mathbf{b}_{,v}(P_2),\;
\mathbf{b}_{,w}(P_2)]
\end{aligned} \; ,
\]

Therefore, the time-dependent global nodal coordinates, as well as the gradients components, $x_{12}$, $y_{12}$, and $z_{12}$,
\begin{align*}
	\mathbf{x}_{12}^\top &\equiv [\, x_1,\; x_{1,u},\; x_{1,v},\; x_{1,w},\; x_2,\; x_{2,u},\; x_{2,v},\; x_{2,w} \,], \\
	\mathbf{y}_{12}^\top &\equiv [\, y_1,\; y_{1,u},\; y_{1,v},\; y_{1,w},\; y_2,\; y_{2,u},\; y_{2,v},\; y_{2,w} \,], \\
	\mathbf{z}_{12}^\top &\equiv [\, z_1,\; z_{1,u},\; z_{1,v},\; z_{1,w},\; z_2,\; z_{2,u},\; z_{2,v},\; z_{2,w} \,] \; .
\end{align*}

as well as the unknown coefficients $\alpha$, $\beta$, and $\gamma$:

\begin{align*}
	\mathbf{B}_{12}^\top \cdot \boldsymbol{\alpha} &= \mathbf{x}_{12}
	\qquad \Rightarrow \qquad
	\boldsymbol{\alpha} = \mathbf{B}_{12}^{-\top} \cdot \mathbf{x}_{12} \; , \\
	\mathbf{B}_{12}^\top \cdot \boldsymbol{\beta}  &= \mathbf{y}_{12}
	\qquad \Rightarrow \qquad
	\boldsymbol{\beta} = \mathbf{B}_{12}^{-\top} \cdot \mathbf{y}_{12} \; , \\
	\mathbf{B}_{12}^\top \cdot \boldsymbol{\gamma} &= \mathbf{z}_{12}
	\qquad \Rightarrow \qquad
	\boldsymbol{\gamma} = \mathbf{B}_{12}^{-\top} \cdot \mathbf{z}_{12} \; .
\end{align*}

which leads to:
\[
[\boldsymbol{\alpha} \;\; \boldsymbol{\beta} \;\; \boldsymbol{\gamma}]
=
\mathbf{B}_{12}^{-\top}
[\mathbf{x}_{12} \;\; \mathbf{y}_{12} \;\; \mathbf{z}_{12}] \; .
\]

,
\[
\renewcommand{\arraystretch}{0.9} 
\left[
\begin{array}{c}
	\boldsymbol{\alpha}^\top \\
	\boldsymbol{\beta}^\top \\
	\boldsymbol{\gamma}^\top
\end{array}
\right]
=
\left[
\begin{array}{c}
	\mathbf{x}_{12}^\top \\
	\mathbf{y}_{12}^\top \\
	\mathbf{z}_{12}^\top
\end{array}
\right]
\mathbf{B}_{12}^{-1} \; ,
\]

which means that:
\[
\renewcommand{\arraystretch}{0.9}
\mathbf{r}(u,v,w,t) =
\left[
\begin{array}{c}
	\mathbf{x}_{12}^\top \\
	\mathbf{y}_{12}^\top \\
	\mathbf{z}_{12}^\top
\end{array}
\right]
\mathbf{B}_{12}^{-1} \cdot \mathbf{b}(u,v,w)
= \mathbf{N}(t) \cdot \mathbf{s}(u,v,w) \; ,
\]

where \( \mathbf{N}(t) \in \mathbb{R}^{3 \times 8} \) and \( \mathbf{s}(u,v,w) \in \mathbb{R}^{8 \times 1} \) are defined as:
\begin{subequations}
	\label{subeq:NU-SF}
	\begin{equation}
		\label{eq:nodal-unknowns-def}
		\mathbf{N}(t) \equiv
		{\renewcommand{\arraystretch}{0.9}
			\left[
			\begin{array}{cccccccc}
				x_1 & x_{1,u} & x_{1,v} & x_{1,w} & x_2 & x_{2,u} & x_{2,v} & x_{2,w} \\
				y_1 & y_{1,u} & y_{1,v} & y_{1,w} & y_2 & y_{2,u} & y_{2,v} & y_{2,w} \\
				z_1 & z_{1,u} & z_{1,v} & z_{1,w} & z_2 & z_{2,u} & z_{2,v} & z_{2,w}
			\end{array}
			\right]
		}
		\notag
	\end{equation}
	\begin{equation}
		\label{equ:formulation-3243-e}
		\begin{aligned}
		= [\, &\mathbf{r}_1 \;\; \mathbf{r}_{1,u} \;\; \mathbf{r}_{1,v} \;\; \mathbf{r}_{1,w} \;\; 
		\mathbf{r}_2 \;\; \mathbf{r}_{2,u} \;\; \mathbf{r}_{2,v} \;\; \mathbf{r}_{2,w} \,] \\
		\equiv [\, &\mathbf{e}_1 \;\; \mathbf{e}_2 \;\; \mathbf{e}_3 \;\; \mathbf{e}_4 \;\; 
		\mathbf{e}_5 \;\; \mathbf{e}_6 \;\; \mathbf{e}_7 \;\; \mathbf{e}_8 \,] \; ,
		\end{aligned}
	\end{equation}
	with \( \mathbf{e}_i \in \mathbb{R}^3 \), for \( 1 \leq i \leq 8 \), and the shape function is defined as
	\begin{equation}
		\label{eq:shape-func-def}
		\mathbf{s}(u,v,w) \equiv \mathbf{B}_{12}^{-1} \cdot \mathbf{b}(u,v,w) \in \mathbb{R}^8 \; .
	\end{equation}
\end{subequations}

\subsection{ANCF 3443 Shell Element}

\begin{figure}[htb]
    \centering
    \safeincludegraphics[width=0.7\textwidth]{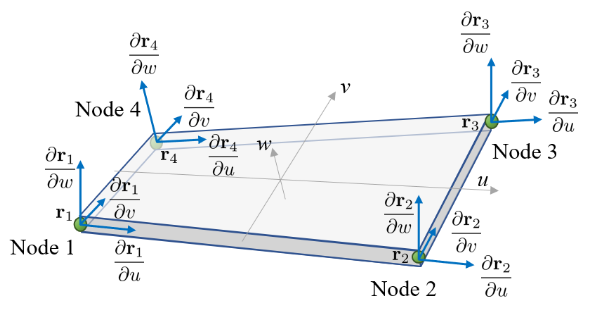}
    \caption{Visualization of a fully-parameterized 4-node ANCF shell element 3443}
    \label{fig:3443shell}
\end{figure}

The 3443 ANCF shell element is a fully parameterized 4-node element, as shown in Fig.~\ref{fig:3443shell}. Each of the four nodes in this quadrilateral element is defined by a position vector and three corresponding position vector gradients. Altogether, this configuration yields 48 degrees of freedom for a single unconstrained element, which is double that of the fully parameterized 2-node ANCF beam element 3243.

The vector of basis function for 3443 shell element is:
\[
\begin{aligned}
\mathbf{b}^T(u,v,w) = [1, u, v, w, uw, vw, uv, u^2, w^2, u^3, w^3, u^2 w, w^2 u, uvw, u^3 w, uw^3]
\end{aligned}
\]

The constant interpolation matrix $\mathbf{B} \in \mathbb{R}^{16 \times 16}$ can be constructed by evaluating the basis functions and their partial derivatives with respect to the material coordinates $(u,v,w)$ at each of the four nodes:
\[
\begin{aligned}
\mathbf{B_{1234}} = [&\mathbf{b}(P_1),\;
\mathbf{b}_{,u}(P_1),\;
\mathbf{b}_{,v}(P_1),\;
\mathbf{b}_{,w}(P_1), \\
&\ldots,\;
\mathbf{b}(P_4),\;
\mathbf{b}_{,u}(P_4),\;
\mathbf{b}_{,v}(P_4),\;
\mathbf{b}_{,w}(P_4)]
\end{aligned}
\]

The time-dependent global nodal coordinates can be grouped similarly to the beam case, into three vectors containing the $x$, $y$, and $z$ components across all nodes and their gradients:
\begin{align*}
	\mathbf{x}_{1234}^\top &\equiv [\, x_1,\; x_{1,u},\; x_{1,v},\; x_{1,w},\;
	x_2,\; x_{2,u},\; x_{2,v},\; x_{2,w},\; \\
	&\phantom{\equiv [} x_3,\; x_{3,u},\; x_{3,v},\; x_{3,w},\;
	x_4,\; x_{4,u},\; x_{4,v},\; x_{4,w} \,], \\
	\mathbf{y}_{1234}^\top &\equiv [\, y_1,\; y_{1,u},\; y_{1,v},\; y_{1,w},\;
	y_2,\; y_{2,u},\; y_{2,v},\; y_{2,w},\; \\
	&\phantom{\equiv [} y_3,\; y_{3,u},\; y_{3,v},\; y_{3,w},\;
	y_4,\; y_{4,u},\; y_{4,v},\; y_{4,w} \,], \\
	\mathbf{z}_{1234}^\top &\equiv [\, z_1,\; z_{1,u},\; z_{1,v},\; z_{1,w},\;
	z_2,\; z_{2,u},\; z_{2,v},\; z_{2,w},\; \\
	&\phantom{\equiv [} z_3,\; z_{3,u},\; z_{3,v},\; z_{3,w},\;
	z_4,\; z_{4,u},\; z_{4,v},\; z_{4,w} \,] \; .
\end{align*}

As with the 3243 element, the unknown coefficient vectors $\boldsymbol{\alpha}$, $\boldsymbol{\beta}$, and $\boldsymbol{\gamma}$ can be determined using the inverse of the matrix $\mathbf{B}$:

\begin{align*}
	\boldsymbol{\alpha} &= \mathbf{B}^{-T} \cdot \mathbf{x}_{1234}, \\
	\boldsymbol{\beta}  &= \mathbf{B}^{-T} \cdot \mathbf{y}_{1234}, \\
	\boldsymbol{\gamma} &= \mathbf{B}^{-T} \cdot \mathbf{z}_{1234} \; .
\end{align*}

Therefore, the position vector field can be expressed as:
\[
\renewcommand{\arraystretch}{0.9}
\mathbf{r}(u,v,w;t) =
\left[
\begin{array}{c}
		\mathbf{x}_{1234}^\top \\
		\mathbf{y}_{1234}^\top \\
	\mathbf{z}_{1234}^\top
\end{array}
\right]
\mathbf{B_{1234}}^{-1} \cdot \mathbf{b}(u,v,w)
= \mathbf{N}(t) \cdot \mathbf{s}(u,v,w) \; ,
\]

where \( \mathbf{N}(t) \in \mathbb{R}^{3 \times 16} \) and \( \mathbf{s}(u,v,w) \in \mathbb{R}^{16 \times 1} \) are defined as:
\begin{subequations}
		\label{subeq:N-sf-3443}
			\begin{equation}
				\mathbf{N}(t) \equiv
			{\renewcommand{\arraystretch}{0.9}\setlength{\arraycolsep}{2pt}%
				\scalebox{0.85}{\ensuremath{\left[
					\begin{array}{cccccccccccccccc}
						x_1 & x_{1,u} & x_{1,v} & x_{1,w} & x_2 & x_{2,u} & x_{2,v} & x_{2,w} & x_3 & x_{3,u} & x_{3,v} & x_{3,w} & x_4 & x_{4,u} & x_{4,v} & x_{4,w} \\
						y_1 & y_{1,u} & y_{1,v} & y_{1,w} & y_2 & y_{2,u} & y_{2,v} & y_{2,w} & y_3 & y_{3,u} & y_{3,v} & y_{3,w} & y_4 & y_{4,u} & y_{4,v} & y_{4,w} \\
						z_1 & z_{1,u} & z_{1,v} & z_{1,w} & z_2 & z_{2,u} & z_{2,v} & z_{2,w} & z_3 & z_{3,u} & z_{3,v} & z_{3,w} & z_4 & z_{4,u} & z_{4,v} & z_{4,w}
					\end{array}
					\right]}}
				}
				\notag
		\end{equation}
	\begin{equation}
		= [\, \mathbf{e}_1 \;\; \mathbf{e}_2 \;\; \dots \;\; \mathbf{e}_{16} \,] \; ,
		\quad \text{with } \mathbf{e}_i \in \mathbb{R}^3
	\end{equation}
	\begin{equation}
		\mathbf{s}(u,v,w) \equiv \mathbf{B_{1234}}^{-1} \cdot \mathbf{b}(u,v,w) \in \mathbb{R}^{16} \; .
	\end{equation}
\end{subequations}

\section{Joint Constraint Details}
\label{sec:joint_details}
This appendix records the lower-level details behind the dot-product constraints introduced in \cref{subsubsec:joints}. The accumulated revolute-joint Jacobian blocks, their insertion into the augmented-Lagrangian residual, and the main consequences for the Newton system are now discussed in the main text. What remains here is the detailed scaling argument behind row normalization and the explicit block form of the DP1 Hessian.

\subsection{Scaling Basis for Constraint Row Normalization}
\label{subsec:rev_normalization}

A mixed CD/DP1 constraint set introduces a magnitude mismatch that can severely degrade Newton convergence. CD constraint values are $\mathcal O(\|\Delta\mathbf r\|)$, whereas DP1 values scale as $\mathcal O(\delta^2)$ when the direction vectors are defined by material fibers of length $\delta$, chosen as a small fraction of the element diameter. The Jacobian entries inherit the same scaling: $\mathcal O(1)$ for CD and $\mathcal O(\delta)$ for DP1.

Two consequences follow. First, the raw Newton residual barely corrects angular errors because the DP1 rows are much smaller than the CD rows. Second, in the Gauss--Newton term $h^2\rho\,\mathbf C_q^{\mathsf T}\mathbf C_q$, the angular eigenvalues are $\mathcal O(\rho\delta^2)$, whereas the translational eigenvalues are $\mathcal O(\rho)$. The condition number is therefore amplified by a factor of order $1/\delta^2$. This is a conditioning artifact introduced by the choice of short material fibers used to define the joint directions; it is not a property of the underlying mechanics.

The remedy adopted in \cref{eq:row_weights,eq:rev_normalized} is per-row normalization with constant weights computed once from the reference configuration. For the revolute joint, the three CD rows are left unscaled, while each DP1 row is divided by the reference magnitude of the two participating directions. When $|\mathbf a_0|=|\mathbf b_{j,0}|=\delta$, the DP1 weights are of order $1/\delta$, so the normalized DP1 rows become $\mathcal O(1)$ at the level of both residual and Jacobian. After normalization, the translational and angular parts of the Gauss--Newton term are both $\mathcal O(\rho)$, which restores a balanced Newton system.

The same argument extends to any joint containing dot-product constraints. The precise weights may differ from one joint type to another, but the principle is unchanged: the constraint rows should be scaled so that their Jacobian magnitudes are commensurate before entering the Newton solve.

\subsection{Explicit DP1 Hessian Blocks}
\label{subsec:rev_hessian}

For the revolute joint of \cref{subsubsec:joints}, let the point set be $(\mathcal P,\mathcal Q)$ on body~$b$ with hosting elements $(E,F)$ and $(\mathcal R,\mathcal S,\mathcal T)$ on body~$c$ with hosting elements $(G,H,K)$. The evaluation operators are those introduced in \cref{eq:eval_op}. Consider first the DP1 constraint
\begin{equation}
  \label{eq:dp1_hessian_c1}
  c_{\mathrm{DP1},1}=\mathbf a^{\mathsf T}\mathbf b_1-f_1,
\end{equation}
with $\mathbf a=\mathbf r_{\mathcal Q}-\mathbf r_{\mathcal P}$ and $\mathbf b_1=\mathbf r_{\mathcal S}-\mathbf r_{\mathcal R}$. Since $c_{\mathrm{DP1},1}$ is a product of two functions that are linear in $\mathbf q$, its Hessian is constant. Ordered as $(\mathbf e_E^b,\mathbf e_F^b,\mathbf e_G^c,\mathbf e_H^c)$, the nonzero block structure is
\begin{equation}
  \label{eq:dp1_hessian}
  \nabla_q^2 c_{\mathrm{DP1},1}
  =
  \begin{bmatrix}
    \mathbf 0 & \mathbf 0 & +\boldsymbol\Sigma_E^{\mathsf T}\boldsymbol\Sigma_G & -\boldsymbol\Sigma_E^{\mathsf T}\boldsymbol\Sigma_H \\[4pt]
    \mathbf 0 & \mathbf 0 & -\boldsymbol\Sigma_F^{\mathsf T}\boldsymbol\Sigma_G & +\boldsymbol\Sigma_F^{\mathsf T}\boldsymbol\Sigma_H \\[4pt]
    +\boldsymbol\Sigma_G^{\mathsf T}\boldsymbol\Sigma_E & -\boldsymbol\Sigma_G^{\mathsf T}\boldsymbol\Sigma_F & \mathbf 0 & \mathbf 0 \\[4pt]
    -\boldsymbol\Sigma_H^{\mathsf T}\boldsymbol\Sigma_E & +\boldsymbol\Sigma_H^{\mathsf T}\boldsymbol\Sigma_F & \mathbf 0 & \mathbf 0
  \end{bmatrix}.
\end{equation}
For the second DP1 row, $c_{\mathrm{DP1},2}=\mathbf a^{\mathsf T}\mathbf b_2-f_2$, the structure is identical, with $\boldsymbol\Sigma_H$ replaced by $\boldsymbol\Sigma_K$.

Three properties are worth recording. First, the Hessian is \emph{constant}: it depends only on shape-function evaluations at the chosen constraint points and can therefore be precomputed once the joint topology is fixed. Second, it is \emph{sparse}: only cross-body couplings appear, while same-body diagonal blocks vanish. Third, it is \emph{indefinite}: the vanishing diagonal blocks together with the nonzero off-diagonal couplings imply eigenvalues of both signs.

These properties explain the structure of the Newton matrix in \cref{eq:newton_hessian,eq:full_newton}. The Gauss--Newton contribution is positive semidefinite, whereas the curvature contribution associated with DP1 rows is indefinite and weighted by the effective multipliers $\eta_i$. Consequently, once the consistent DP1 Hessian is retained, the full Newton system is symmetric but not, in general, positive definite. This is the algebraic reason that dot-product constraints require a symmetric-indefinite solve if one wishes to preserve the consistent quadratic-convergence model.

\section{Select Material Models}
\label{sec:select_material_models}
\subsection{Hyperelastic Material Laws}

\subsubsection{St Venant-Kirchhoff (SVK) material}
The Saint-Venant--Kirchhoff (SVK) material model is a fundamental hyperelastic constitutive model in finite element analysis (FEA), particularly suited to the total Lagrangian formulation, where it employs the Green--Lagrange strain tensor and the second Piola--Kirchhoff stress to capture moderate nonlinear deformations in compressible solids~\cite{Bonet2008nonlinear}. This model extends classical linear elasticity to finite strains by defining a quadratic strain energy density function $\Psi(\mathbf{E})$ (see Eq.~\eqref{eq:svk_strain_energy}), with $\lambda$ and $\mu$ denoting the Lam\'e constants, making it computationally straightforward for quasi-static and dynamic simulations of structures undergoing large rotations but limited stretches~\cite{Bonet2022limitations}. Popular in computational mechanics for its simplicity and compatibility with isoparametric elements, the SVK model finds broad use in beam and shell analyses, such as Timoshenko--Ehrenfest beams, biological soft tissues, and multifunctional laminates, although its popularity wanes in extreme compression due to unphysical softening beyond about $58\%$ strain~\cite{Zur2023large}. Despite its limitations in large uniaxial compression, it remains a standard benchmark for validating more advanced hyperelastic models, such as neo-Hookean or Mooney--Rivlin formulations, in FEA software like Abaqus~\cite{Luo2020benchmark}.

The strain energy density function is quadratic in $\mathbf{E}$:
\begin{equation}
  \Psi(\mathbf{E})
  = \frac{\lambda}{2} \bigl(\operatorname{tr}(\mathbf{E})\bigr)^2
    + \mu \, \operatorname{tr}\bigl(\mathbf{E}^2\bigr) ,
  \label{eq:svk_strain_energy}
\end{equation}
where $\lambda$ and $\mu$ are the Lam\'e constants.

The second Piola--Kirchhoff stress follows from differentiation:
\begin{equation}
  \mathbf{S}
  = \lambda \, \operatorname{tr}(\mathbf{E}) \, \mathbf{I}
    + 2 \mu \, \mathbf{E} .
  \label{eq:svk_second_piola}
\end{equation}

From this, the first Piola--Kirchhoff stress is:
\begin{equation}
  \mathbf{P}
  = \mathbf{F}\mathbf{S}
  = \lambda \, \operatorname{tr}(\mathbf{E}) \, \mathbf{F}
    + 2 \mu \, \mathbf{F}\mathbf{E} .
  \label{eq:svk_first_piola}
\end{equation}

Using $\mathbf{E} = \tfrac{1}{2}\bigl(\mathbf{F}^T\mathbf{F} - \mathbf{I}\bigr)$,
this can also be written as:
\begin{equation}
  \mathbf{P}
  = \lambda \left(
      \tfrac{1}{2}\operatorname{tr}(\mathbf{F}^T\mathbf{F})
      - \tfrac{3}{2}
    \right)\mathbf{F}
    + \mu \, \mathbf{F}\mathbf{F}^T\mathbf{F}
    - \mu \, \mathbf{F} .
  \label{eq:svk_first_piola_expanded}
\end{equation}

The SVK model is only accurate for small to moderate strains,
although it can accommodate arbitrarily large rotations.
For large strains, it can give non-physical predictions,
such as negative stiffness in compression.

For hyperelastic materials, the first Piola--Kirchhoff stress satisfies
$\mathbf{P} = \partial \Psi/\partial \mathbf{F}$; for the SVK model this is consistent with
Eqs.~\eqref{eq:svk_first_piola} and~\eqref{eq:svk_first_piola_expanded}.

Using the definition of internal force in the Total Lagrangian setting, the internal force associated with
nodal unknown $\mathbf{e}_i$ is, by definition,
\begin{align}
\mathbf{f}_i^{T}
  &:= \frac{\partial}{\partial \mathbf{e}_i}
      \int_{V_r} \Psi(\mathbf{E}) \,\mathrm{d}V_r
   = \int_{V_r} \frac{\partial \Psi(\mathbf{E})}{\partial \mathbf{e}_i}\,
      \mathrm{d}V_r \\
  &= \int_{V_r} \frac{\partial \Psi(\mathbf{E})}{\partial \mathbf{F}}
      : \frac{\partial \mathbf{F}}{\partial \mathbf{e}_i} \,\mathrm{d}V_r
   = \int_{V_r} \mathbf{P}
      : \frac{\partial \mathbf{F}}{\partial \mathbf{e}_i} \,\mathrm{d}V_r \\
  &= \int_{V_r} (\mathbf{P}\mathbf{h}_i)^{T} \,\mathrm{d}V_r .
\end{align}
which leads to the following result:
\begin{equation}
\mathbf{f}_i = \int_{V_r} \mathbf{P}\,\mathbf{h}_i\,\mathrm{d}V_r ,
\label{eq:hyperelastic_internal_force_def}
\end{equation}
where $\mathbf{P}$ is the first Piola--Kirchhoff stress tensor and
$\mathbf{h}_i := s_{i,\mathbf{u}}^{T} \in \mathbb{R}^{3\times 1}$ is the reference-gradient of the shape function $s_i(u,v,w)$.
Note that this important relation holds for \emph{any} hyperelastic
material model, not only for the SVK material model.
Plugging in the expression of $\mathbf{P}$ for the SVK material model,
see Eq.~\eqref{eq:svk_first_piola_expanded}, we get:
\begin{equation}
\mathbf{f}_i
  = \int_{V_r}
      \Bigl[
        \lambda \Bigl( \tfrac{1}{2}\operatorname{tr}(\mathbf{F}^{T}\mathbf{F}) - \tfrac{3}{2} \Bigr)\mathbf{F}
        + \mu \mathbf{F}\mathbf{F}^{T}\mathbf{F}
        - \mu \mathbf{F}
      \Bigr]\mathbf{h}_i \,\mathrm{d}V_r ,
\label{eq:svk_internal_force}
\end{equation}
and therefore, the expression of the virtual work of the internal force
over the element is given by:
\begin{equation}
\delta W_{\text{int}}
  = \sum_{i=1}^{n_u} \delta \mathbf{e}_i^{T} \cdot \mathbf{f}_i .
\label{eq:internal_virtual_work}
\end{equation}

Second-order (Newton-type) methods require the consistent linearization
of the internal force vector, i.e., the Jacobian (tangent stiffness) $\partial \mathbf{f}_i/\partial \mathbf{e}_j$.
 For the SVK material model, this Jacobian reads
 \begin{equation}
 \begin{aligned}
 \frac{\partial \mathbf{f}_i}{\partial \mathbf{e}_j}
   &= \int_{V_r}
       \Bigl[
         \lambda (\mathbf{F}\mathbf{h}_i)(\mathbf{F}\mathbf{h}_j)^{T}
         + \lambda\operatorname{tr}(\mathbf{E})\,(\mathbf{h}_j^{T}\mathbf{h}_i)\,\mathbf{I} \\
   &\quad + \mu (\mathbf{F}\mathbf{h}_j)^{T}(\mathbf{F}\mathbf{h}_i)\,\mathbf{I}
          + \mu (\mathbf{F}\mathbf{h}_j)(\mathbf{F}\mathbf{h}_i)^{T} \\
   &\quad + \mu (\mathbf{h}_j^{T}\mathbf{h}_i)\,\mathbf{F}\mathbf{F}^{T}
          - \mu (\mathbf{h}_j^{T}\mathbf{h}_i)\,\mathbf{I}
       \Bigr] \,\mathrm{d}V_r .
 \end{aligned}
 \label{eq:svk_tangent}
 \end{equation}

\subsubsection{Mooney-Rivlin (MR) material}

The Mooney--Rivlin material model is a widely adopted hyperelastic constitutive framework for finite element analysis (FEA) of rubber-like materials, formulated in terms of strain invariants \(I_1\) and \(I_2\) of the right Cauchy--Green tensor to accommodate the near-incompressible behavior of elastomers~\cite{Mooney1940theory,Rivlin1948large}. In this work, we employ the compressible two-parameter form in Eq.~\eqref{eq:mr_strain_energy}, which augments the isochoric Mooney--Rivlin response with a volumetric penalty term~\cite{Bonet2008nonlinear}. The model is effective for moderate to large strains and is widely used for elastomers and soft tissues due to its computational efficiency and compatibility with total Lagrangian formulations~\cite{Luo2020benchmark}.

It is expressed directly in terms of the deformation gradient $\mathbf{F}$
and the invariants of the right Cauchy--Green tensor
$\mathbf{C} = \mathbf{F}^T \mathbf{F}$. It depends on three material parameters: $\mu_{10}$ and $\mu_{01}$ (shear response) and $k$ (bulk modulus controlling compressibility).
The Mooney--Rivlin model reduces to the neo-Hookean model when $\mu_{01} = 0$,
and to a purely incompressible form when $k \to \infty$ with $J = 1$.

The compressible Mooney--Rivlin strain energy density function is given by
\begin{equation}
\Psi(\mathbf{F})
= \mu_{10}\,(\bar{I}_1 - 3) + \mu_{01}\,(\bar{I}_2 - 3) + \frac{k}{2}\,(J - 1)^2 ,
  \label{eq:mr_strain_energy}
\end{equation}
where
\begin{align}
  \mathbf{C} &= \mathbf{F}^T \mathbf{F}
  &&\text{(right Cauchy--Green tensor)}, \label{eq:mr_C_def} \\
  I_1 &= \operatorname{tr}(\mathbf{C}), \label{eq:mr_I1_def} \\
  I_2 &= \tfrac{1}{2}\bigl[(\operatorname{tr}\mathbf{C})^2
          - \operatorname{tr}(\mathbf{C}^2)\bigr], \label{eq:mr_I2_def} \\
  J &= \det \mathbf{F}, \label{eq:mr_J_def} \\
  \bar{I}_1 &= J^{-2/3} I_1, \label{eq:mr_I1bar_def} \\
  \bar{I}_2 &= J^{-4/3} I_2. \label{eq:mr_I2bar_def}
\end{align}

The first Piola--Kirchhoff stress can be computed as:
\begin{equation}
  \mathbf{P} = \frac{\partial \Psi}{\partial \mathbf{F}}
\label{eq:mr_first_piola_def}
\end{equation}
using the chain rule. First, the required derivative identities are:
\begin{align}
  \frac{\partial I_1}{\partial \mathbf{F}} &= 2\mathbf{F}, \label{eq:mr_dI1_dF} \\
  \frac{\partial I_2}{\partial \mathbf{F}} &= 2\bigl(I_1\mathbf{F} - \mathbf{F}\mathbf{C}\bigr),
  \label{eq:mr_dI2_dF} \\
  \frac{\partial J}{\partial \mathbf{F}} &= J \mathbf{F}^{-T}, \label{eq:mr_dJ_dF} \\
  \frac{\partial J^{-m}}{\partial \mathbf{F}} &= -m J^{-m} \mathbf{F}^{-T}. \label{eq:mr_dJm_dF}
\end{align}

Therefore, the computation of the derivatives of the isochoric invariants gives:
\begin{align}
  \frac{\partial \bar{I}_1}{\partial \mathbf{F}}
  &= \frac{\partial (J^{-2/3} I_1)}{\partial \mathbf{F}}
   = J^{-2/3} \frac{\partial I_1}{\partial \mathbf{F}}
     + I_1 \frac{\partial J^{-2/3}}{\partial \mathbf{F}} \notag \\
  &= J^{-2/3} (2\mathbf{F})
     - \frac{2}{3} I_1 J^{-2/3} \mathbf{F}^{-T} \notag \\
	  &= 2 J^{-2/3} \mathbf{F}
	     - \frac{2}{3} J^{-2/3} I_1 \mathbf{F}^{-T}.
	  \label{eq:mr_dI1bar_dF}
\end{align}

Similarly,
\begin{equation}
\begin{aligned}
\frac{\partial \bar I_2}{\partial \mathbf{F}}
&= \frac{\partial (J^{-4/3} I_2)}{\partial \mathbf{F}} \\
&= J^{-4/3} \frac{\partial I_2}{\partial \mathbf{F}}
   + I_2 \frac{\partial J^{-4/3}}{\partial \mathbf{F}} \\
&= J^{-4/3} \bigl[ 2 ( I_1 \mathbf{F} - \mathbf{F}\mathbf{C}) \bigr]
   - \frac{4}{3} I_2 J^{-4/3} \mathbf{F}^{-T} \\
&= 2 J^{-4/3} ( I_1 \mathbf{F} - \mathbf{F}\mathbf{C})
   - \frac{4}{3} J^{-4/3} I_2 \mathbf{F}^{-T}.
\end{aligned}
\label{eq:mr_dI2bar_dF}
\end{equation}

The derivative of the volumetric penalty term is:
\begin{equation}
\frac{\partial}{\partial \mathbf{F}}
\left[\frac{k}{2}(J-1)^2\right]
= k(J-1)\frac{\partial J}{\partial \mathbf{F}}
= k(J-1)J\mathbf{F}^{-T}.
\label{eq:mr_dvolumetric_dF}
\end{equation}

After the final assembly based on the above presented equation:
\begin{equation}
\begin{aligned}
\mathbf{P}
&= \mu_{10} \frac{\partial \bar I_1}{\partial \mathbf{F}}
 + \mu_{01} \frac{\partial \bar I_2}{\partial \mathbf{F}}
 + k (J-1) J \mathbf{F}^{-T} \\
&= 2 \mu_{10} J^{-2/3} \mathbf{F}
   - \frac{2}{3} \mu_{10} J^{-2/3} I_1 \mathbf{F}^{-T} \\
&\quad + 2 \mu_{01} J^{-4/3} ( I_1 \mathbf{F} - \mathbf{F}\mathbf{C})
   - \frac{4}{3} \mu_{01} J^{-4/3} I_2 \mathbf{F}^{-T} \\
&\quad + k (J-1) J \mathbf{F}^{-T}.
\end{aligned}
\label{eq:mr_first_piola_assembled}
\end{equation}

Thus, the first Piola--Kirchhoff stress for the compressible
Mooney--Rivlin model is:
\begin{equation}
\begin{aligned}
\mathbf{P}
&= 2 \mu_{10} J^{-2/3}
    \bigl( \mathbf{F} - \tfrac{1}{3} I_1 \mathbf{F}^{-T} \bigr) \\
&\quad + 2 \mu_{01} J^{-4/3}
    \bigl( I_1 \mathbf{F} - \mathbf{F}\mathbf{C}
          - \tfrac{2}{3} I_2 \mathbf{F}^{-T} \bigr) \\
&\quad + k (J-1) J \mathbf{F}^{-T}
\end{aligned}
\label{eq:mr_first_piola}
\end{equation}

Using Eq.~\eqref{eq:hyperelastic_internal_force_def} together with Eq.~\eqref{eq:mr_first_piola}, the internal force associated with the nodal unknown
$\mathbf{e}_i$ admits the decomposition
\begingroup
\normalsize
\begin{equation}
\begin{aligned}
\mathbf{f}_i
  &= \mathbf{t}_1 - \mathbf{t}_2 + \mathbf{t}_3 - \mathbf{t}_4 - \mathbf{t}_5 + \mathbf{t}_6, \\
\mathbf{t}_1 &= 2\mu_{10} \int_{V_r} J^{-2/3} \, \mathbf{F}\mathbf{h}_i \,\mathrm{d}V_r,
&\qquad
\mathbf{t}_4 &= 2\mu_{01} \int_{V_r} J^{-4/3} \, \mathbf{F}\mathbf{F}^T\mathbf{F}\mathbf{h}_i \,\mathrm{d}V_r, \\
\mathbf{t}_2 &= \frac{2\mu_{10}}{3} \int_{V_r} J^{-2/3} I_1 \, \mathbf{F}^{-T}\mathbf{h}_i \,\mathrm{d}V_r,
&\qquad
\mathbf{t}_5 &= \frac{4\mu_{01}}{3} \int_{V_r} J^{-4/3} I_2 \, \mathbf{F}^{-T}\mathbf{h}_i \,\mathrm{d}V_r, \\
\mathbf{t}_3 &= 2\mu_{01} \int_{V_r} J^{-4/3} I_1 \, \mathbf{F}\mathbf{h}_i \,\mathrm{d}V_r,
&\qquad
\mathbf{t}_6 &= k \int_{V_r} J\,(J-1)\, \mathbf{F}^{-T}\mathbf{h}_i \,\mathrm{d}V_r.
\end{aligned}
\label{eq:mr_internal_force_decomp}
\end{equation}
\endgroup

Second-order (Newton-type) methods require the Jacobian of the internal force,
$\mathbf{K}_{ij} := \partial \mathbf{f}_i/\partial \mathbf{e}_j \in \mathbb{R}^{3\times 3}$.
Differentiating Eq.~\eqref{eq:mr_internal_force_decomp} yields
\begin{equation}
\begin{aligned}
\frac{\partial \mathbf{f}_i}{\partial \mathbf{e}_j}
  &= \frac{\partial \mathbf{t}_1}{\partial \mathbf{e}_j}
   - \frac{\partial \mathbf{t}_2}{\partial \mathbf{e}_j}
   + \frac{\partial \mathbf{t}_3}{\partial \mathbf{e}_j}
   - \frac{\partial \mathbf{t}_4}{\partial \mathbf{e}_j}
   - \frac{\partial \mathbf{t}_5}{\partial \mathbf{e}_j}
   + \frac{\partial \mathbf{t}_6}{\partial \mathbf{e}_j}, \\
\frac{\partial \mathbf{t}_1}{\partial \mathbf{e}_j}
  &= 2\mu_{10} \int_{V_r} J^{-2/3}
      \Bigl[ (\mathbf{h}_j^{T}\mathbf{h}_i)\,\mathbf{I}
             - \frac{2}{3}(\mathbf{F}\mathbf{h}_i)(\mathbf{F}^{-T}\mathbf{h}_j)^{T} \Bigr] \,\mathrm{d}V_r, \\
\frac{\partial \mathbf{t}_2}{\partial \mathbf{e}_j}
  &= \frac{2\mu_{10}}{3} \int_{V_r} J^{-2/3}
      \Bigl[ (\mathbf{F}^{-T}\mathbf{h}_i)(2\mathbf{F}\mathbf{h}_j)^{T} \\
  &\quad - \frac{2}{3} I_1 (\mathbf{F}^{-T}\mathbf{h}_i)(\mathbf{F}^{-T}\mathbf{h}_j)^{T} \\
  &\quad - I_1 (\mathbf{F}^{-T}\mathbf{h}_j)(\mathbf{F}^{-T}\mathbf{h}_i)^{T} \Bigr] \,\mathrm{d}V_r, \\
\frac{\partial \mathbf{t}_3}{\partial \mathbf{e}_j}
  &= 2\mu_{01} \int_{V_r} J^{-4/3}
      \Bigl[ I_1 (\mathbf{h}_j^{T}\mathbf{h}_i)\,\mathbf{I}
             + (\mathbf{F}\mathbf{h}_i)(2\mathbf{F}\mathbf{h}_j)^{T} \\
  &\quad - \frac{4}{3} I_1 (\mathbf{F}\mathbf{h}_i)(\mathbf{F}^{-T}\mathbf{h}_j)^{T} \Bigr] \,\mathrm{d}V_r, \\
\frac{\partial \mathbf{t}_4}{\partial \mathbf{e}_j}
  &= 2\mu_{01} \int_{V_r} J^{-4/3}
      \Bigl[ (\mathbf{F}\mathbf{h}_j)^{T}(\mathbf{F}\mathbf{h}_i)\,\mathbf{I}
             + (\mathbf{F}\mathbf{h}_j)(\mathbf{F}\mathbf{h}_i)^{T} \\
  &\quad + (\mathbf{h}_j^{T}\mathbf{h}_i)\,\mathbf{F}\mathbf{F}^{T}
             - \frac{4}{3}\,\mathbf{F}\mathbf{F}^{T}(\mathbf{F}\mathbf{h}_i)(\mathbf{F}^{-T}\mathbf{h}_j)^{T} \Bigr] \,\mathrm{d}V_r, \\
\frac{\partial \mathbf{t}_5}{\partial \mathbf{e}_j}
  &= \frac{4\mu_{01}}{3} \int_{V_r} J^{-4/3}
      \Bigl[ (\mathbf{F}^{-T}\mathbf{h}_i)\Bigl(
               2(\mathbf{F}\mathbf{h}_j)^{T}(I_1\mathbf{I}-\mathbf{F}\mathbf{F}^{T}) \\
  &\quad - \frac{4}{3} I_2 (\mathbf{F}^{-T}\mathbf{h}_j)^{T}
             \Bigr)
             - I_2 (\mathbf{F}^{-T}\mathbf{h}_j)(\mathbf{F}^{-T}\mathbf{h}_i)^{T} \Bigr] \,\mathrm{d}V_r, \\
\frac{\partial \mathbf{t}_6}{\partial \mathbf{e}_j}
  &= k \int_{V_r} J
      \Bigl[ (2J-1) (\mathbf{F}^{-T}\mathbf{h}_i)(\mathbf{F}^{-T}\mathbf{h}_j)^{T} \\
	  &\quad - (J-1) (\mathbf{F}^{-T}\mathbf{h}_j)(\mathbf{F}^{-T}\mathbf{h}_i)^{T} \Bigr] \,\mathrm{d}V_r.
\end{aligned}
\label{eq:mr_tangent}
\end{equation}

\end{appendices}

\bibliography{BibFiles/refsGraphics,BibFiles/refsSensors,BibFiles/refsAutonomousVehicles,BibFiles/refsChronoSpecific,BibFiles/refsDEM,BibFiles/refsFSI,BibFiles/refsMBS,BibFiles/refsRobotics,BibFiles/refsSBELspecific,BibFiles/refsTerramech,BibFiles/refsCompSci,BibFiles/refsNumericalIntegr,BibFiles/refsMLPhysics,BibFiles/refsSurfaceTension,BibFiles/refsStatsML,BibFiles/refsOddsEnds,BibFiles/refsML-AI,BibFiles/refsLinAlgebra}

\end{document}